\documentclass[preprint]{aastex631}
\pdfoutput=1
\usepackage{natbib}
\usepackage{xcolor}
\shorttitle{Kinematics of SGRE-associated CMEs}
\shortauthors{M{\"a}kel{\"a} et al.}
\graphicspath{{./}{figures/}}

\begin{document}

\title{Speed and Acceleration of CMEs Associated with Sustained Gamma-Ray Emission Events
	Observed by Fermi/LAT}

\correspondingauthor{Pertti M{\"a}kel{\"a}}
\email{pertti.makela@nasa.gov, makela@cua.edu}

\author[0000-0002-0786-7307]{Pertti M{\"a}kel{\"a}}
\affiliation{The Catholic University of America \\
620 Michigan Ave., N.E. \\
 Washington, DC 20064, USA}
\affiliation{NASA Goddard Space Flight Center \\
	8800 Greenbelt Road \\
	Greenbelt, MD 20771, USA}

\author{Nat Gopalswamy}
\affiliation{NASA Goddard Space Flight Center \\
8800 Greenbelt Road \\
Greenbelt, MD 20771, USA}

\author{Sachiko Akiyama}
\affiliation{The Catholic University of America \\
	620 Michigan Ave., N.E. \\
	Washington, DC 20064, USA}
\affiliation{NASA Goddard Space Flight Center \\
	8800 Greenbelt Road \\
	Greenbelt, MD 20771, USA}

\author{Hong Xie}
\affiliation{The Catholic University of America \\
	620 Michigan Ave., N.E. \\
	Washington, DC 20064, USA}
\affiliation{NASA Goddard Space Flight Center \\
	8800 Greenbelt Road \\
	Greenbelt, MD 20771, USA}

\author{Seiji Yashiro}
\affiliation{The Catholic University of America \\
	620 Michigan Ave., N.E. \\
	Washington, DC 20064, USA}
\affiliation{NASA Goddard Space Flight Center \\
	8800 Greenbelt Road \\
	Greenbelt, MD 20771, USA}



\begin{abstract}

The sustained gamma-ray emission (SGRE) from the Sun is a prolonged enhancement of $>$100 MeV gamma-ray emission that extends beyond the flare impulsive phase. The origin of the $>$300 MeV protons resulting in SGRE is debated, both flares and shocks driven by coronal mass ejections (CMEs) being the suggested sites of proton acceleration. We compared the near-Sun acceleration and space speed of CMEs with 'Prompt' and 'Delayed' (SGRE) gamma-ray components \citep{2021ApJS..252...13A}. We found that 'Delayed'-component-associated CMEs have higher initial acceleration and space speed than 'Prompt-only'-component-associated CMEs. We selected halo CMEs (HCMEs) associated with type II radio bursts (shock-driving HCMEs) and compared the average acceleration and space speed between HCME populations with or without SGRE events, major solar energetic particle (SEP) events, metric, or decameter-hectometric (DH) type II radio bursts. We found that the SGRE-producing HCMEs associated with a DH type II radio burst and/or a major SEP event have higher {space speeds and especially initial accelerations} than those without an SGRE event. We estimated the radial distance and speed of the CME-driven shocks at the end time of the 2012 January 23 and March 07 SGRE events using white-light images of STEREO Heliospheric Imagers and radio dynamic spectra of Wind WAVES. The shocks were at the radial distances of 0.6--0.8 au and their speeds were high enough ($\approx$975 km s$^{-1}$ and $\approx$750 km s$^{-1}$, respectively) for high-energy particle acceleration. Therefore, we conclude that our findings support the CME-driven shock as the source of $>$300 MeV protons.

\end{abstract}



\section{Introduction} \label{sec:intro}

The sustained gamma-ray emission (SGRE) from the Sun is a prolonged enhancement of $>$100 MeV gamma-ray emission that extends beyond the flare impulsive phase.  SGRE typically lasts for several hours, extending well beyond the end of the associated soft X-ray flare emission. The first SGRE event at energies above 100 MeV was detected on 1991 June 15 by the Gamma-1 telescope on board the Gamma spacecraft and it lasted at least 2.16 hours \citep{1991ICRC....3...73A}. Similar observation of a long-duration $>$50 MeV gamma-ray emission was reported by \cite{1993A&AS...97..349K} during the 1991 June 11 flare. The $>$100 MeV SGRE is produced by $>$300 MeV protons precipitating from the solar corona into the solar chromosphere, where their interactions with the dense plasma layers create pions, which then decay into the observed $>$100 MeV gamma-rays \citep[e.g.,][]{1987ApJS...63..721M}. The dominant source of $>$100 MeV gamma-rays is neutral pion decay \citep[e.g.,][]{2018ApJ...864..148K,2020Atoms...8...14G}. \cite{1985ICRC....4..146F} first reported a clear detection of $>$40 MeV gamma-rays that require pion production during the extended phase of the 1982 June 3 gamma-ray flare.
SGRE events were originally called long duration gamma-ray flares \citep[LDGRFs; e.g.,][]{2000SSRv...93..581R}. Nowadays they are also known as late-phase $>$100 MeV gamma-ray emission \citep[LPGRE; e.g.,][]{2018ApJ...869..182S} events. A review of gamma-ray observations analogous to Gamma-1 measurements by \cite{2000SSRv...93..581R} listed 13 LDGRFs between 1982--1991. A few more early events have been discovered from observations by non-dedicated gamma-ray telescopes \citep{2014ASSL..400..301K}. Most recently, observations by the Large Area Telescope \citep[LAT,][]{2009ApJ...697.1071A} on board the Fermi satellite have shown that SGRE events are relatively common \citep{2014ApJ...787...15A,2017ApJ...835..219A,2014ApJ...789...20A,2021ApJS..252...13A,2015ApJ...805L..15P,2018ApJ...869..182S, 2018_allafort_thesis}. The $>$100 MeV SGRE event on 2012 March 7 was observed to last over 20 hours \citep[e.g.,][]{2018ApJ...868L..19G,2018ApJ...869..182S}.
 
The origin of the $>$300 MeV protons producing SGRE is still debated. \cite{1987ApJS...63..721M} studied the 1982 June 03 event and suggested a two-phase particle acceleration scenario, where a short-duration impulsive-phase acceleration is followed by a second acceleration phase, probably due protons accelerated by coronal shocks and resulting in SGREs \citep[see also,][]{1987ApJ...316L..41R}. \cite{1991ApJ...368..316R} investigated the same 1983 June 03 gamma-ray flare and found a good agreement between their model of turbulent solar flare loops and the observed gamma-ray light-curves, including the extended emission phase, which their model explained to be due to delayed protons diffusing both in momentum space and spatially in the flare loops. The flare loop scenario requires that flare-accelerated protons must remain trapped and/or be continuously re-accelerated in the coronal loops long after the X-ray flare itself has ended. However, trapping of high-energy protons in coronal loops for several hours requires force-free loops {\citep{1993ApJ...414..908L}} with a sufficiently low density and turbulence level \citep{2000SSRv...93..581R}. As an alternative to particle trapping in the coronal loops, \cite{1994SoPh..150..267K} suggested a continuous stochastic acceleration due to additional pulses of energy that could explain gamma-ray observations during the extended phase of the 1991 June 15 LDGRF \citep[see also,][]{1991ICRC....3...73A,1996SoPh..166..107A}.

Gamma-ray-line observations of the behind-the-limb flare on 1989 September 29 were interpreted to require a spatially extended gamma-ray source and hence to suggest shocks driven by fast and wide coronal mass ejections (CMEs) as a likely source of the gamma-ray-emission producing particles \citep[e.g.,][]{1993ICRC....3...91C,1993ApJ...409L..69V}. Recent LAT observations of SGRE events during eruptions occurring behind the solar limb, have confirmed that an extended source of gamma-rays must exist at the Sun \citep[e.g.,][]{2015ApJ...805L..15P,2017A&A...608A..43P}. The CME-driven shock naturally extends over large regions of solar surface allowing the shock-accelerated protons to have access to areas far from the behind-the-limb eruption site. \cite{2020SoPh..295...18G} forward modelled the CME flux rope and the surrounding shock in the 2014 September 1 behind-the-limb event and found that the Fermi/LAT SGRE source was located far from the flare site - in the space between the flux rope and shock confirming the extended nature of the emission. \cite{2018IAUS..335...49H} suggested another scenario where closed magnetic loops extended up to the height of several solar radii will capture high-energy protons that might be accelerated by a CME shock and subsequently the loops retract and enable sufficiently large number of $>$300 MeV protons to interact with the solar atmosphere. 

Recently, \cite{2019ApJ...879...90D} compared the estimated fluxes of gamma-ray producing particles precipitating into the solar atmosphere with the fluxes of SEPs escaping into interplanetary space and did not find significant correlation. They suggested that the lack of correlation rules out the CME-driven shock as a common source of both fluxes. However, \cite{2021ApJ...915...82G} pointed out that the correlation is high when systematic effects are corrected differently. \cite{2020ApJ...893...76K} compared the SGRE time profile observed during the ground level enhancement (GLE) on 2017 September 10 with the time profiles of simulated shock parameters and found a good match between them, supporting the CME shock as a common source of SGRE-producing protons at the Sun and GLE protons at 1 au. \cite{2018ApJ...864...39W} studied properties of flares and CMEs with and without SGREs. They found that SGRE events are associated with intense X-class flares but only one-third of the X-class solar flares Fermi/LAT observed have an SGRE event. They also note that fast and wide CMEs are associated with SGRE events. Therefore, their results on the flare and CME associations favor CME-driven shock as the source of $>$300 MeV protons.

\begin{figure}[ht!]
\plotone{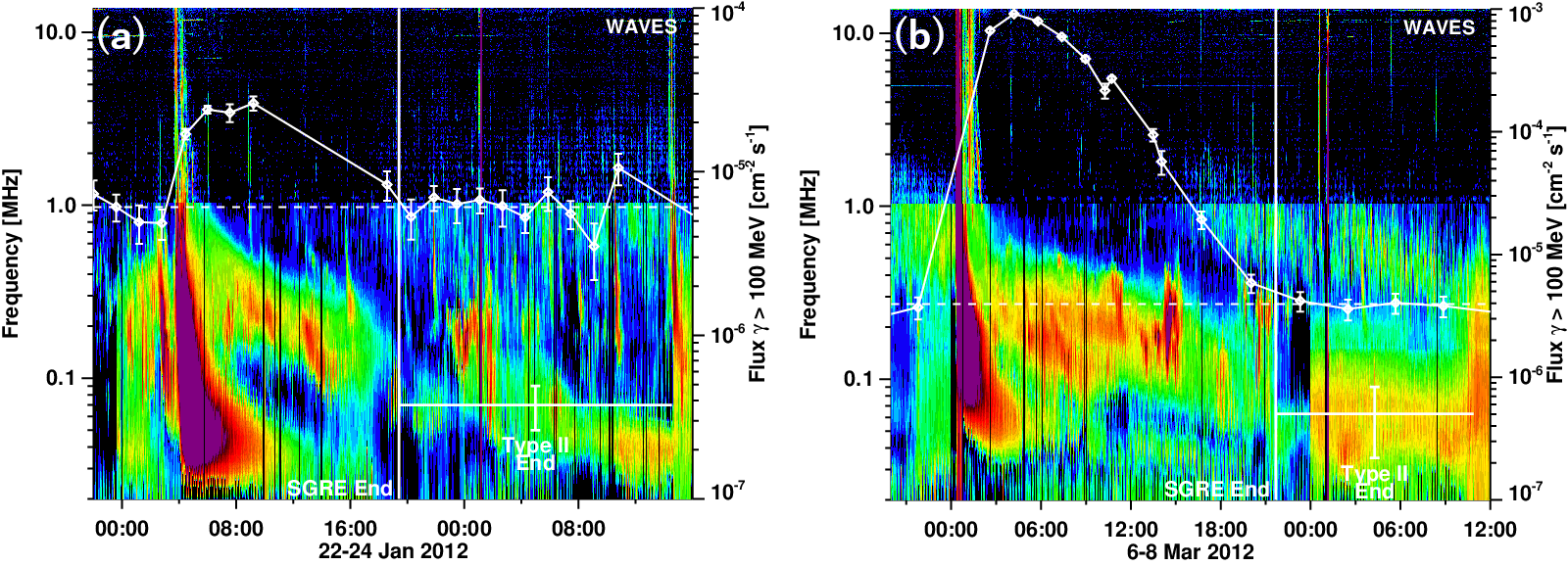}
\caption{Data points of Fermi/LAT $>$100 MeV gamma-ray flux overplot on Wind/WAVES radio dynamic spectra for the 2012 January 23 and March 07 SGRE events. The LAT data points are connected by straight lines to guide the eye. The estimated end times of the SGRE events and type II radio bursts are marked by long and short vertical lines.  \label{fig:events}}
\end{figure}

Additional support for the CME-shock scenario is provided by the correlation of the SGRE durations with the durations and the end frequencies of type II radio bursts \citep{2018ApJ...868L..19G,2019JPhCS1332a2004G}. Figure \ref{fig:events} shows two examples of concurrent SGRE events and type II radio bursts during SGRE event in January and March 2012. Although type II radio bursts are produced by CME-shock accelerated electrons, they indicate the presence of a strong shock that could also accelerate protons to high energies. Therefore, the correlations suggest that CME-driven shocks could be the source of both the electrons resulting in the decameter-hectometric (DH) type II radio bursts and the $>$300 MeV protons generating the SGRE events. \cite{2022ApJ...929..172P} investigated the EUV wave connection to the behind-the-limb (BTL) flare at S20E140 on 2021 July 17. They found that the time when the EUV wave crosses the limb onto the visible disk  and the onset of the LAT $>$100 MeV flux enhancement are concurrent. They also found a coupling between the peak times of the time derivative of the EUV wave intensity profile observed at 193 {\AA} and the $>$100 MeV gamma-ray flux suggesting that the EUV wave and the acceleration of the SGRE-producing protons are connected. They found the correlation to be valid in three other Fermi/LAT BTL flares. \cite{2022ApJ...929..172P} conclude that the correlation between the derivative of the EUV wave intensity and gamma-ray flux and the near-simultaneous appearance of  a complex type II radio burst indicate that radio, EUV and gamma-ray emissions share the same source (CME-shock) although the emissions originate at different heights in the corona.

Back-precipitation of shock-accelerated protons have been studied using numerical simulations but results so far have not been accordant with one another. \cite{2022A&A...658A..23H} modelled particle precipitation including enhanced turbulence and found scattering to increase back-precipitation but even that being the case the fraction of protons able to precipitate down to the radial distance of 1 R$_\sun$ relative to the injected back-propagating protons is less than 1\%. The precipitation fraction decreases as a function of the radial distance of the CME shock. Therefore, they conclude that the CME-driven shocks cannot provide a sufficient flux of $>$300 MeV protons to explain the SGRE events. Opposite conclusions in support of a CME-shock as the source of the gamma-ray-producing protons have been obtained by \cite{2018ApJ...867..122J} who studied the Fermi behind-the-limb flare on 2014 September 1. Their simulations of the CME-driven shock indicated that the quasi-perpendicular part of the shock had a magnetic connection to the gamma-ray source at the front-side of the Sun and the shock compression ratio increase matched the increase in the observed gamma-ray emission. \cite{2018ASSL..444..157A} simulated proton acceleration in the CME-driven shocks during the 2012 January 23 and May 17 SGRE events. The 2012 May 17 SGRE event was also observed as a GLE by neutron monitors. They concluded that proton acceleration by coronal shocks and diffusive downstream particle transport could explain the SGRE events. However, the authors of the above-mentioned studies suggest that more elaborated MHD models for the particle transport back to the Sun is required because the complex structure of the magnetic fields near the Sun, which the current simulation efforts cannot fully replicate. The lack of direct observations of the precipitating protons close to the Sun leaves the question whether they can propagate back to the solar atmosphere deep enough open. 

The initial acceleration and speed of the CME in part control the formation height and strength of the shock, which in turn affect particle acceleration efficiency of the shock. Therefore, the CME acceleration and speed provide a proxy for the effectiveness of high-energy particle acceleration in the CME-driven shocks. \cite{2022CEAB..46...1G} studied SGRE association with on-disk CMEs producing major SEP events and HCMEs with sky-speeds $\ge$1800 km s$^{-1}$ during cycle 24. They investigated the initial acceleration and space speed of the CMEs, which they defined to be the instantaneous peak space speed and acceleration obtained from forward fitting of the graduated cylindrical shell (GCS) flux rope model \citep{2009SoPh..256..111T,2011ApJS..194...33T} to the EUV and coronagraph images of the CMEs. They found that the peak space speed and peak initial acceleration of the SGRE-producing CME are 2516 km s$^{-1}$ and 3.87 km s$^{-2}$, respectively. \cite{2022CEAB..46...1G} suggest that the close connection they found between CME kinematics and the SGRE events give support to the CME-shock scenario.

In addition to SEP events, type II radio burst are related to particle acceleration by CME-driven shocks. In this report we estimate the initial acceleration and space speed of the CMEs associated with the Fermi/LAT solar flares (FLSFs) during solar cycle 24 listed by \cite{2021ApJS..252...13A}. In order to evaluate the feasibility of the CME-driven shocks in producing SGRE events, we compare average initial acceleration and space speed of CME populations associated with SGRE and SEP events and type II radio bursts. We use space speeds obtained by applying geometrical correction to close-to-the-limb CMEs or by applying the model by \cite{2004JGRA..109.3109X} to HCMEs. Initial acceleration is estimated by assuming that the CME obtains its estimated space speed during the interval extending from the onset time to the peak time of the associated soft X-ray flare \citep{2012SSRv..171...23G, 2015ApJ...806...13M}. In addition, we estimate the radial distance and the space speed of shocks at the end time of the two longest-duration SGRE events on 2012 January 23 and March 07.

\section{Data} \label{sec:data}

In the analysis we use the catalog published by \cite{2021ApJS..252...13A} that contains 45 FLSFs with $>$30 MeV gamma-ray emission in the period 2010 January--2018 January. We do not repeat here all the details of the event data analysis, which are given in \cite{2021ApJS..252...13A}. We briefly describe their categorization method of FLSFs. \cite{2021ApJS..252...13A} characterized the light curves of the FLSFs based on the associated hard X-ray (HXR) observations made by the Fermi Gamma-ray Burst Monitor \citep[GBM,][]{2009ApJ...702..791M}. If the early evolution of the gamma-ray emission was synchronous with the Fermi/GBM HXR evolution, the flare was deemed to have an impulsive 'Prompt' component lasting $\lesssim$10 minutes. If the flare had a second phase of gamma-ray emission without a corresponding HXR evolution, the flare was deemed to have a gradual 'Delayed' component that could last up to $\approx$20 hours. \cite{2021ApJS..252...13A} found that a total of 39 out of the 45 FLSFs had detectable level of $>$100 MeV emission. One should note that Fermi/LAT does not observe the Sun continuously, the average LAT measurement interval lasts about 30 minutes \citep{2021ApJS..252...13A}. Of those 45 FLSFs, they classified 6 flares as 'Prompt only' and 4 flares as 'Delayed only'. In 10 flares both the 'Prompt' and 'Delayed' emission were detected by LAT and 6 flares were detected with LAT Low Energy (LLE) analysis only.

The existence of the DH type II radio bursts is based on Wind spacecraft's radio and plasma wave instrument \citep[WAVES,][]{1995SSRv...71..231B} observations (\url{https://cdaw.gsfc.nasa.gov/CME_list/radio/waves_type2.html}){, STEREO/WAVES instrument \citep{2008SSRv..136..487B} observations,} and on the analysis by {\cite{2018ApJ...868L..19G,2019JPhCS1332a2004G}}. The metric type II radio burst and  soft X-ray flare observations are obtained from the NOAA Solar and Geophysical Event Reports. We adjusted the NOAA-reported flare onset times in some events after inspecting concurrent EUV images and soft X-ray curves of the solar eruption. The CME data near the Sun is provided by the Large Angle and Spectrometric Coronagraph \citep[LASCO,][]{1995SoPh..162..357B} on the Solar and Heliospheric Observatory \citep[SOHO,][]{1995SoPh..162....1D} spacecraft.  The CME data is collected from the SOHO/LASCO CME Catalogs (\url{https://cdaw.gsfc.nasa.gov/CME_list/index.html}, \url{https://cdaw.gsfc.nasa.gov/CME_list/halo/halo.html}). SEP event data are from the Major SEP Event list (\url{https://cdaw.gsfc.nasa.gov/CME_list/sepe/}) {and from the GOES-equivalent $>$10 MeV intensities calculated using data provided by the High Energy Telescope \citep[HET;][]{2008SSRv..136..391V} onboard STEREO}. For the shock distance estimation at the end of the SGRE event, we used white-light images of the Sun Earth Connection Coronal and Heliospheric Investigation \citep[SECCHI,][]{2008SSRv..136...67H} Heliospheric Imagers \citep[HIs,][]{2009SoPh..254..387E} onboard the Solar Terrestrial Relations Observatory \citep[STEREO,][]{2008SSRv..136...17D} spacecraft. The HI images were provided by the STEREO Archive maintained by the UK Solar System Data Centre (\url{https://www.ukssdc.ac.uk/solar/stereo/data.html}). To identify the associated CMEs, we inspected the CME catalogues provided by the Heliospheric Cataloguing, Analysis and Techniques Service (HELCATS, \url{https://www.helcats-fp7.eu/}).

\section{Estimation Method of the CME Initial Acceleration} \label{sec:estim}

The initial acceleration of the CME near the Sun is difficult to measure because the cadence of white-light coronagraphs is limited. In our study we follower the method previously used by \cite{2012SSRv..171...23G} and \cite{2015ApJ...806...13M}. We assume that the CME accelerates from rest to its final maximum speed, which it reaches at the peak time of the associated soft X-ray flare. \cite{2006ApJ...649.1100Z} have shown that the main acceleration phase of the CME coincides with the impulsive phase of the associated X-ray flare. Therefore, we calculate the initial acceleration a of the CME with a formula: $a=V_{Space}/(t_{FlarePeak}-t_{FlareOnset})$, where $V_{Space}$ is the estimated space speed of the CME and $t_{FlarePeak}$ and $t_{FlareOnset}$ are the flare peak and onset times, respectively. The space speed of halo CMEs (HCMEs) has been estimated by using a cone model for HCMEs \citep{2004JGRA..109.3109X} and the space speeds are listed in the SOHO/LASCO HALO CME catalog (\url{https://cdaw.gsfc.nasa.gov/CME_list/halo/halo.html}). For non-HCMEs, the space speed, $V_{Space}$, is calculated from the measured CME speed on the sky plane, $V_{Sky}$, by using a geometrical correction $V_{Sky}/\cos{\theta}$, where $\theta$ is the angle the CME propagation direction makes away from the sky plane. The angle $\theta$ depends on the longitude of the flare location. To avoid unrealistically large corrections, we have included in the analysis only non-HCMEs for which $\theta$ is $\leq$30$^{\circ}$ as seen either from the SOHO or STEREO spacecraft. The method calculates an average over the acceleration phase of the CME. The peak initial acceleration of the CME can be higher than obtained average initial acceleration as was shown by \cite{2021ApJ...915...82G}.

In general, we know that the CME speed profiles near the Sun vary from event to event and CME speed is an important parameter governing particle acceleration efficiency of the CME-driven shocks. \cite{2016ApJ...833..216G,2017JPhCS.900a2009G} showed that CMEs associated with major SEP events have a hierarchical relationship between the initial acceleration and speed of the CME and the SEP fluence spectral indices \citep[see also,][]{2019SunGe..14..123M}: CMEs associated with filament eruptions have low initial speeds and acceleration and produce the softest SEP spectra at 1 au, while the CMEs with highest initial speed and acceleration have the hardest SEP spectra. The CMEs with an intermediate speed and acceleration result in moderately hard SEP spectra at 1 au. Therefore, initial acceleration and speed provide a proxy for the effectiveness of high-energy particle acceleration in the CME-driven shocks. 

\subsection{Initial Acceleration and Space Speed of CMEs Associated with LAT Gamma-ray Flares} \label{subsec:init}

In our analysis we use the on-disk gamma-ray events listed in \cite{2021ApJS..252...13A}. Their list contains 45 gamma-ray flares during cycle 24. \cite{2021ApJS..252...13A} categorized the flares based on whether 'Prompt' or 'Delayed' component (SGRE event) of gamma-ray emission was detected. In the 6 events of the 45 events, only 'Prompt' (impulsive) emission was detected, 4 events had no detected 'Prompt' emission at all, 10 events have both 'Prompt' and 'Delayed' emission and the remaining 25 had 'Delayed' emission, but the presence of 'Prompt' emission could not be excluded because of LAT was not pointing to the Sun at the appropriate time. 32 of the flares were associated with a HCME, 10 were associated with non-HCMEs, and 3 had no associated CME. Based on our own estimations, we changed the CME of the 2014 September 10 flare to the 08:00 UT HCME. We have excluded the 3 back sided flares, the 3 flares without a CME and the 2017 September 06 X2.2 flare for which we could not estimate the space speed of the CME at 09:48 UT, because there is no suitable side-view either from SOHO or STEREO-A.

Table \ref{tab:tbl1} list the total number and the average value of the initial acceleration and space speed of CMEs in different categories. First, we divided the CMEs into two main categories: those associated with flares showing only a 'Prompt' component, labelled as 'Prompt Only' and those with a 'Delayed' component, labelled 'All Delayed' in Table \ref{tab:tbl1}. The 'Prompt Only' flares are impulsive gamma-ray flares and the 'All Delayed' ones are SGRE events. Clearly, the impulsive gamma-ray flares are associated with significantly slower CMEs (775 km s$^{-1}$) than flares with an SGRE event (1708 km s$^{-1}$). The difference in the initial acceleration is not as clear, but again the CMEs with SGRE events show a larger initial acceleration than those without an SGRE event. From SEP event comparisons by \citeauthor{2016ApJ...833..216G} (\citeyear{2016ApJ...833..216G}; see also \citealt{2019SunGe..14..123M}) we know that higher acceleration and speed indicate that the CME-driven shock produces harder energy spectra, i.e., more likely to have $>$300 MeV protons. Similar high initial acceleration and fast speed characteristics are shared by CMEs associated with GLEs, which are guaranteed to have $>$300 MeV protons.

Then we divided the 'All Delayed' CMEs into three subcategories: the 'Prompt Delayed' CMEs are associated with gamma-ray flares having both emission components, the 'No-Prompt Delayed' CMEs do not have a detectable 'Prompt' component and the 'Delayed' CMEs have a 'Delayed' component but the existence of the 'Prompt' component is uncertain because of the lack of LAT observations during the impulsive phase of the flare. Differences are now less significant (the sample sizes also become small), but the 'Prompt Delayed' CMEs appear to have the highest average initial acceleration and space speed and the 'No-prompt Delayed' the lowest ones among the three groups. Most likely the CMEs without associated 'Prompt' gamma-ray component are more slowly accelerating CMEs but are still able to produce $>$300 MeV protons as their space speed becomes high enough in the later phase. Again, similar slower initial acceleration but high later-phase speed has been detected for CMEs producing major SEP events \citep{2016ApJ...833..216G}. Table \ref{tab:tbl1_data} in Appendix lists the data for events included in calculations of Table \ref{tab:tbl1}.

\begin{deluxetable*}{lccccccc}
\tabletypesize{\scriptsize}
\tablewidth{0pt} 
\tablenum{1}
\tablecaption{Initial acceleration and space speed of CMEs associated with LAT gamma-ray flares \label{tab:tbl1}}
\tablehead{
\colhead{Quantity} & \colhead{} & \multicolumn{2}{c}{Main Types} & \colhead{} & \multicolumn{3}{c}{Subtypes of 'All Delayed'}  \\
\cline{3-4}
\cline{6-8}
\colhead{} & \colhead{} &
\colhead{Prompt Only} & \colhead{All Delayed} & \colhead{} & \colhead{Delayed} &
\colhead{Prompt Delayed} & \colhead{No-Prompt Delayed} \\
\colhead{(1)} & \colhead{} & \colhead{(2)} & \colhead{(3)} & \colhead{} & \colhead{(4)} &
\colhead{(5)} & \colhead{(6)}
} 
\startdata 
{Count                     } && 6 & 32 && 18 & 8 & 4 \\
{Mean Acceleration (km s$^{-2}$)} && 1.37 & 1.75 && 1.73 & 1.87 & 1.62 \\
{Mean Space Speed (km s$^{-1}$) } && 775 & 1708 && 1745 & 1753 & 1663 \\
\enddata
\end{deluxetable*}

\section{Comparison with HCMEs Associated with Type II Radio Bursts and Major SEP Events} \label{sec:comp}

Because the 'All Delayed' gamma-ray flares are mainly associated with HCMEs, we compare their initial acceleration and space speed with HCMEs associated with type II radio bursts and major SEP events. Major SEP events are defined as those with the peak proton flux in the GOES $>$10 MeV integral channel above 10 particles cm$^{-2}$ s$^{-1}$ sr$^{-1}$. Since SEPs are charged particles, they spiral along the interplanetary magnetic field lines as they propagate away from the acceleration source. Therefore, at Earth we can detect mostly SEP events originating from eruptions occurring in the western hemisphere of the Sun. Some very intense eruptions from the eastern limb can produce particle events at Earth but in that case only at the lower energies.

In general, DH type II radio bursts are well correlated with major SEP events \citep{2002ApJ...572L.103G,2019JPhCS1332a2004G,2004ApJ...605..902C}. Both radio and gamma-ray emissions can be detected from all on-disk eruptions because electromagnetic emission can propagate away from the Sun without being significantly affected by coronal or interplanetary medium. Type II solar radio bursts occur at the fundamental and second harmonic of local plasma frequency that depends on the electron density at the upstream of the CME shock.  Because the electron number density decreases as a function of the radial distance, the plasma frequency decreases away from the Sun and higher frequency emissions originating from a lower height can propagate freely outwards. Therefore, the type II burst can be identified in the radio dynamic spectra as an intensity feature slowly drifting towards lower frequencies at the rate that depends on the shock speed  and the density scale height of the ambient medium. 
\begin{deluxetable*}{lcccccccc}
\tabletypesize{\scriptsize}
\tablewidth{0pt} 
\tablenum{2}
\tablecaption{Initial acceleration and space speed of cycle-24 HCMEs with metric type II radio bursts \label{tab:tbl2}}
\tablehead{
\colhead{HCME Category} & \colhead{} & \multicolumn{3}{c}{'Delayed' Component (SGRE Event)} & \colhead{} & \multicolumn{3}{c}{No 'Delayed' Component}  \\
\cline{3-5}
\cline{7-9}
\colhead{} & \colhead{} &
\colhead{Count} & \colhead{Mean Acceleration} & \colhead{Mean Space Speed} & \colhead{} & \colhead{Count} &
\colhead{Mean Acceleration} & \colhead{Mean Space Speed} \\
\colhead{} & \colhead{} & \colhead{} & \colhead{(km s$^{-2}$)} & \colhead{(km s$^{-1}$)} & \colhead{} &
\colhead{} & \colhead{(km s$^{-2}$)} & \colhead{(km s$^{-1}$)} \\
\colhead{(1)} & \colhead{} & \colhead{(2)} & \colhead{(3)} & \colhead{(4)} & \colhead{} & \colhead{(5)} &
\colhead{(6)} & \colhead{(7)}
} 
\startdata 
{DH Type II             } && 23 & 1.85 & 1869 && 37 & 1.09 & 1211 \\
{No DH Type II          } && 1 & \multicolumn{2}{c}{Too low statistics} && 26 & 1.11 & 959 \\
{SEP Event              } && 17 & 1.83 & 2004 && 13 & 1.15 & 1499 \\
{SEP Event (w/STEREO)   } && 21 & 1.70 & 1858 && 18 & 1.07 & 1396 \\
{No SEP Event           } && 7 & 1.81 & 1360 && 50 & 1.09 & 1006 \\
{No SEP Event (w/STEREO)} && 3 & 2.68 & 1524 && 45 & 1.11 &  992 \\
\enddata
\end{deluxetable*}

In Table \ref{tab:tbl2} we have divided the 87 cycle-24 HCMEs with metric type II radio emission into CMEs with and without a 'Delayed' gamma-ray component. The existence of the metric type II radio burst indicates that a shock forms early, making these HCMEs good candidates for SGRE production. We investigated how many of the metric type II-associated HCMEs are with and without a DH type II radio burst or a major SEP event. {We selected major SEP events as they are intense events and could have enhancements of $>$300 MeV protons, which are unlikely to be present in the inherently low-intensity SEP events.} Because observer's connection to the SEP source affects the possibility to detect SEPs, the group without a major SEP event could still contain events that were able to accelerate particles, especially the poorly connected eastern hemisphere events could have produced high-energy particles that were not detected. {We account for this possibility by using GOES $>$10 MeV equivalent STEREO intensities to identify major SEP events observed by STEREO. The STEREO $>$10 MeV flux is estimated using data from the STEREO/HET \citep{2008SSRv..136..391V}, which covers the energy ranges of 13--100 MeV. The flux is estimated by fitting a power law to HET data points and integrating the flux in the 10--150 MeV range \citep[see][]{2016ApJ...833..216G}. In Tables \ref{tab:tbl2} and \ref{tab:tbl3}, we have separated the two SEP event sets and marked the one containing both GOES and STEREO events as "(w/STEREO)", although in Table \ref{tab:tbl3} the statistics for the SGRE events are mostly too low. One should note that STEREO spacecraft drift around the Sun, so their magnetic connection to the Sun changes continuously. In addition, STEREO-A observations have significant data gaps during solar conjunction period during 2014--2015 and contact to STEREO-B was lost on October 1, 2014. We surveyed also STEREO/WAVES data for additional DH type II radio bursts but we found only one on 03 August 2011. The STEREO-A data showed a short-duration, slanted feature in the 10--14 MHz frequency range starting at 13:38 UT, which we added to our DH type II burst list. All other STEREO/WAVES DH type II bursts were accompanied with a Wind/WAVES DH type II burst, so we study STEREO and Wind DH type II bursts together.} The most western event of the 7 SGRE events without a major GOES SEP event occurred at the heliographic longitude W18 and 4 of the 7 SGRE events occurred less than 30$\degr$ from the eastern limb. {The two bottom rows of Table \ref{tab:tbl2} are} difficult to interpret, but we have added {them} mainly for completeness. Results show that all HCMEs associated with an SGRE event have similar average initial acceleration values {(1.70--1.85 km s$^{-2}$) with the exception of the group without a GOES or STEREO SEP event, which contains only three events and the average initial acceleration (2.68 km s$^{-2}$) is very high, possibly indicating missed major SEP event identification. The range is considerably higher than those of HCMEs without an SGRE event (1.07--1.15 km s$^{-2}$). The SGRE and SEP-associated HCMEs have the highest average space speed, whereas two groups of HCMEs, SGRE-associated HCMEs without an SEP event and SEP-associated HCMEs without an SGRE event, seem to have similar speeds. However, the average initial accelerations of SGRE-associated HCMEs without a major SEP event are higher (even when we ignore the group without a GOES or STEREO SEP event that has only 3 events in total) than those of without SGRE event but with a major SEP event.} The DH type II-associated HCMEs without an SGRE event have only slightly lower average space speed ({1211} km s$^{-1}$), but we know that DH type II bursts are associated with SEP events and this mixed population includes 12 HCMEs with an SEP event, which have high space speeds. If we exclude these 12 SEP-associated events, then the average space speed of the remaining {25} events decreases down to {1064} km s$^{-1}$. Clearly, the existence of $>$300 MeV protons is connected to a high initial acceleration and speed of the associated HCME. {The HCMEs without an SEP and SGRE event have the lowest average speed (992 km s$^{-2}$). Therefore, the SGRE-associated HCMEs conform to the hierarchy between the initial acceleration and speed of the CME and the fluence spectral index as described by \cite{2016ApJ...833..216G}. Especially the high initial acceleration seems to be crucial for SGRE production.}

The average accelerations and speeds of the 20 cycle-24 HCMEs associate with only a DH type II radio burst are shown in Table \ref{tab:tbl3}. HCMEs are mostly without SGRE events (only 4 SGRE events) or major SEP events {(only 6 SEP events if STEREO observations are included, two of which have also an SGRE event). Therefore, statistics for SGRE events are low, but the average space speed of SGRE events without an SEP event ($\approx$1579 km s$^{-1}$) is below the space speed of the SGRE events with an SEP event ($\approx$2004 km s$^{-1}$; $\approx$1858 km s$^{-1}$ if STEREO observations are included}) in Table \ref{tab:tbl2}. None of the three eruptions without an major SEP event detected by GOES were magnetically well-connected to Earth, so they probably accelerated high-energy particles efficiently but particles didn't reach Earth. {One of them, the 10 June 2014 HCME with a solar source at S17E82, actually had a major SEP event observed by STEREO-B, which was located at the heliographic longitude E164. The 05 March 2012 HCME had a solar source at N17E52, but the GOES-equivalent $>$10 MeV intensities observed by STEREO-B at longitude E117 were already elevated above 100 pfu due to a preceding HCME on 04 March that was not associated with an SGRE event. At the onset of the 10 March 2012 HCME launched from N17W24, the $>$10 MeV intensities were elevated above 10 pfu at all three spacecraft. In fact, the March 5 and 10 events are the first and last events in a cluster of 4 SGRE events accompanied by high level of SEP flux \citep{2019JPhCS1332a2004G}. So, it is quite possible that the two March 2012 events also accelerated particles.} The average initial acceleration value is lower than respective value for SGRE events with SEP events in Table \ref{tab:tbl2}, but this is expected because CMEs associated with only a DH type II radio burst accelerate slowly and the shock forms later. This probably explains the lower average space speed near the Sun. The initial acceleration and average space speed of HCMEs without an SGRE event and an SEP event are lower or similar, respectively, to the respective values in Table \ref{tab:tbl2}. Table \ref{tab:tbl2_data} in Appendix lists the data for events included in calculations of Tables \ref{tab:tbl2} and \ref{tab:tbl3}.

\begin{deluxetable*}{lcccccccc}
\tabletypesize{\scriptsize}
\tablewidth{0pt} 
\tablenum{3}
\tablecaption{Initial acceleration and space speed of cycle-24 HCMEs with DH type II radio bursts only \label{tab:tbl3}}
\tablehead{
\colhead{HCME Category} & \colhead{} & \multicolumn{3}{c}{'Delayed' Component (SGRE Event)} & \colhead{} & \multicolumn{3}{c}{No 'Delayed' Component}  \\
\cline{3-5}
\cline{7-9}
\colhead{} & \colhead{} &
\colhead{Count} & \colhead{Mean Acceleration} & \colhead{Mean Space Speed} & \colhead{} & \colhead{Count} &
\colhead{Mean Acceleration} & \colhead{Mean Space Speed} \\
\colhead{} & \colhead{} & \colhead{} & \colhead{(km s$^{-2}$)} & \colhead{(km s$^{-1}$)} & \colhead{} &
\colhead{} & \colhead{(km s$^{-2}$)} & \colhead{(km s$^{-1}$)} \\
\colhead{(1)} & \colhead{} & \colhead{(2)} & \colhead{(3)} & \colhead{(4)} & \colhead{} & \colhead{(5)} &
\colhead{(6)} & \colhead{(7)}
} 
\startdata 
{SEP Event} && 1 & \multicolumn{2}{c}{Too low statistics} && 3 & 0.46 & 1263 \\
{SEP Event (w/STEREO)} && 2 & \multicolumn{2}{c}{Too low statistics} && 4 & 0.38 & 1265 \\
{No SEP Event } && 3 & 1.00 & 1579 && 13 & 0.44 & 1164 \\
{No SEP Event (w/STEREO)} && 2 & \multicolumn{2}{c}{Too low statistics} && 12 & 0.47 & 1155 \\
\enddata
\end{deluxetable*}

\section{Radial Distance of the Shock at the End of the SGRE Events} \label{sec:enddist}

We selected two SGRE events, the 2012 January 23 and March 07 events, with the longest duration of the associated type II radio burst (Gopalswamy et al. 2019) and for which the STEREO observations provided sideview white-light images of the HCMEs.

The 2012 January 23 04:00 UT HCME produced a DH type II burst with a duration about {25.0$\pm$9.6} hr, while the estimated duration of the SGRE event was {15.4$\pm$0.8} hr. The SGRE ended around 19:25 UT. The estimated space speed was 2511 km s$^{-1}$, and the interplanetary shock arrived at the SOHO spacecraft at 14:33 UT on January 24. The eruption was associated with a M8.7 X-ray flare starting at 03:38 UT at the heliographic location of N28W21. The STEREO-A and STEREO-B longitudes were W108 and E114, respectively. The eruption produced a major SEP event at Earth with a GOES $>$10 MeV peak proton flux 6310 cm$^{-2}$ s$^{-1}$ sr$^{-1}$. The 2012 January 23 eruption close to the Sun has been studied extensively because the eruption involved two flux ropes that merged below the radial distance of 15 R$_\sun$ \citep[e.g.][]{2013ApJ...769L..25C,2013AdSpR..52....1J,2013A&A...552L..11L}.

The second HCME at 00:24 UT on 2012 March 07 was associated with an SGRE event that had even longer duration, about {21.3$\pm$1.6} hr \citep{2014ApJ...789...20A,2018ApJ...868L..19G}. The 2012 March 07 SGRE had a slightly longer estimated duration of about 21 hr but the SGRE durations cannot be measured accurately because LAT does not observe the Sun continuously. The SGRE end time was 21:40 UT. The estimated duration of the DH type was {27.9$\pm$6.8} hr.  The LASCO space speed of the HCME was 3146 km s$^{-1}$ and it was associated with a X5.4 X-ray flare at 00:02 UT from N17E27. A second X1.3-class flare started about an hour later at 01:05 UT. The associated HCME at 01:30 UT had a slightly slower space speed of 2160 km s$^{-1}$. The STEREO-A and STEREO-B longitudes were W109 and E118, respectively. The SOHO shock arrival time was at 10:53 UT on March 08. The GOES $>$10 MeV peak proton flux was 6530 cm$^{-2}$ s$^{-1}$ sr$^{-1}$. The onset of the HCME has been studied by \cite{2015ApJ...809...34C} and the heliopsheric propagation by \cite{2016ApJ...817...14P} and \cite{2023FrASS...949906S}.

\subsection{Distance Estimation}

We estimated the radial distance and the space speed of the shock by forward fitting a spheroidal shock model to white-light images of STEREO/HIs \citep{2009SoPh..254..387E} around the end time of the SGRE event. For shock fitting, we used IDL programs in the Solar Corona Ray-Tracing Software package developed to forward modeling of structures of the solar corona \citep[e.g.,][\url{https://soho.nascom.nasa.gov/solarsoft/stereo/secchi/idl/scraytrace/}]{2009SoPh..256..111T}.

\begin{figure}[ht!]
\plotone{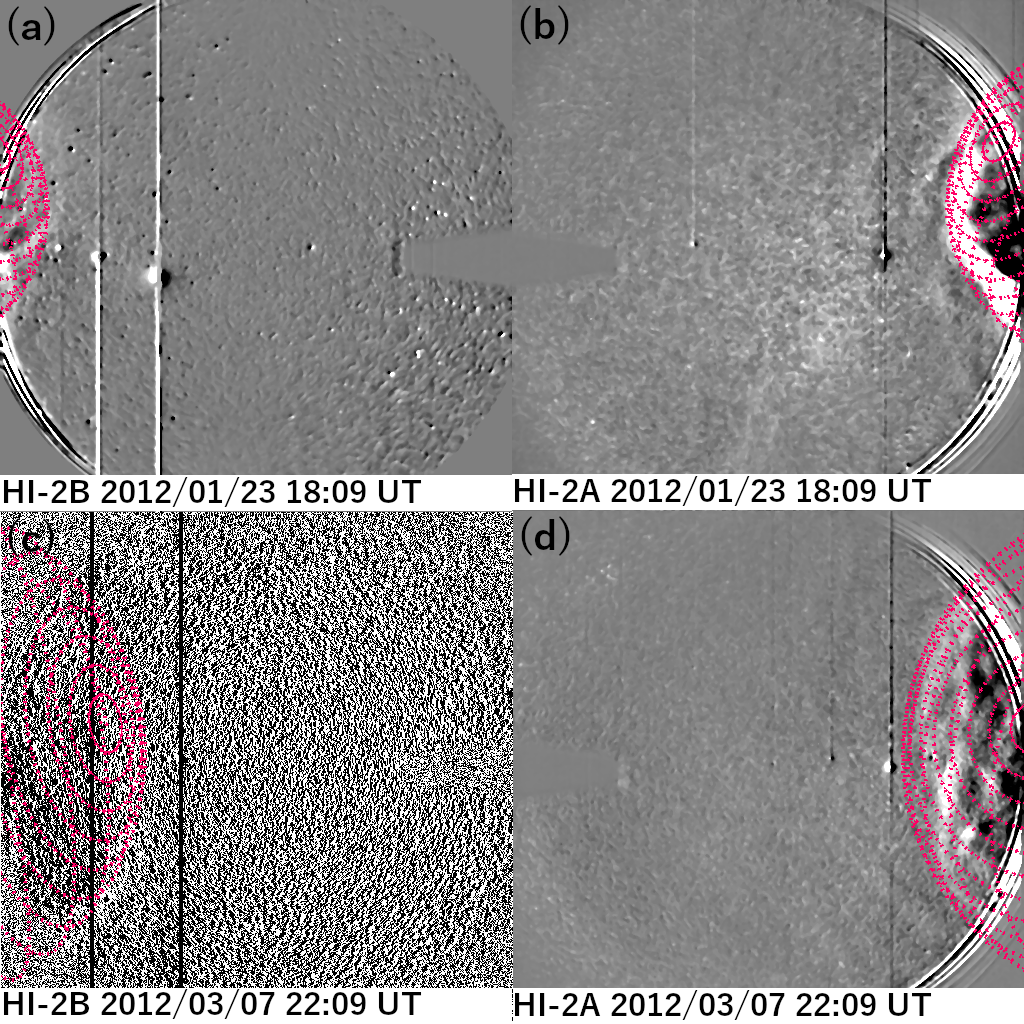}
\caption{Spheroidal shock fits (red lines) on to the STEREO/HI images of the 2012 January 23 CME (top row) and the 2012 March 07 CME (bottom row).  \label{fig:shockfit}}
\end{figure}

The fitting of the spheroidal shock model to HI observations is shown in Figure \ref{fig:shockfit}. The propagation direction of the shock is difficult to estimate, our estimates were N25W05 for the 2012 January 23 CME and N34E27 for the 2012 March 07 CME. For the 2012 January 23 HCME we obtained the radial distance r=121 R$_\sun$, and the space speed 975 km s$^{-1}$. In the case of the 2012 March 07 HCME the estimated radial r=140 R$_\sun$, and the space speed 750 km s$^{-1}$. The obtained speeds are reasonably high for a strong CME-driven shock to exist.

We compared these results with radial distances estimated using Wind/WAVES observations of the type II radio burst. First, we measured the mid-frequency of type II emission lane at the time the CME leading edge was around 20 R$_\sun$, because type II emissions are often very complex and overlapped by more intense type III emission during the early phase of the eruption, which makes radio measurements at frequencies corresponding the shock distances close to the Sun difficult. From the frequency formula $f_{plasma}=9.0 \times \sqrt{N \times n(r)}$, where the radial distance $r$ is in units of R$_\sun$ and frequency $f$ in kHz, we calculated the multiplier $N$ for the Leblanc density model $n(r)$ \citep{1998SoPh..183..165L}. The measurement time was obtained by extrapolating the CME height-time profiles obtained by forward fitting a flux rope model to LASCO and SECCHI/COR images to a radial distance 20 R$_\sun$.

We then estimated the radial distance at the SGRE end time from the mid-frequency of the type II emission lane: For the 2012 January 23 HCME we obtained the multiplier $N= 4.51$, which then gave for the mid-frequency $f=83$ kHz the radial distance of $r=132$ R$_\sun$. For the 2012 March 07 HCME the respective values were $N=9.07$, $f=90$ kHz and $r=173$ R$_\sun$. The distances estimated from the radio bursts data are 9\% and 24\% larger than those estimated from the STEREO/HI images. The STEREO/HI height-time measurements are complicated because the actual shape, location, and propagation direction of the shock ahead of the CME body are difficult to discern from the white-light images. The CME structure in white-light is also transparent, so we may confuse structures \citep[e.g.,][]{2019SpWea..17..539S} and brightness depends on local density and Thomson-scattering geometry \citep{2009SSRv..147...31H,2013SoPh..285..369X}. On the other hand, type II radio emissions are sporadic and depend on local density at the radio source, which the general density model cannot capture. We also assume that the location of the radio source is at the shock nose \citep[e.g.,][]{2018ApJ...867...40M} and the type II emission in interplanetary space occurs at the fundamental of the plasma frequency \citep{1985A&A...151..215L}.

\section{Discussion}

In the first part of our analysis, we showed that the near-Sun kinematics of the CMEs correlate with the properties of the gamma-ray emission observed by Fermi/LAT. The population of the CMEs (total of 8 CMEs) that were associated with a gamma-ray event whose light-curve indicated both 'Prompt' and 'Delayed' emission component, as defined by \cite{2021ApJS..252...13A}, had the highest average initial acceleration (1.87 km s$^{-2}$) and fastest average space speed (1753 km s$^{-1}$). The mixed 'Delayed' category, where the existence of the 'Prompt' component is uncertain due to the lack of LAT measurements around the flare onset, has a similar average space speed (1745 km s$^{-1}$) but somewhat lower initial acceleration (1.73 km s$^{-2}$). The lowest average values (1.37 km s$^{-2}$ and 775 km s$^{-1}$, respectively) had the population of CMEs (total of 6 CMEs) associated with gamma-ray flares showing a 'Prompt' emission component only, i.e., there were no SGRE emissions detected by LAT. The speeds correspond well with those obtained by \cite{2018ApJ...864...39W} who studied CME properties for X-class flares with and without gamma-ray emission. They found a median CME linear speed of 768 km s$^{-1}$ for X-class flares without gamma-ray emission. If Fermi detected gamma-rays during the X-class flare, the median speed of the associated CMEs was 1828 km s$^{-1}$. CMEs associated with SGRE events had the highest median speed of 2125 km s$^{-1}$. The definition of SGRE in their study was that the $>$100 MeV gamma-ray duration is $\gtrsim\sim2$ hr. \cite{2021ApJS..252...13A} definition used here is based on details of hard X-ray and gamma-ray light-curves, which probably explains why \cite{2018ApJ...864...39W} SGRE events were associated with faster CMEs. 

In addition, we divided cycle 24 on-disk HCMEs associated with type II radio bursts into groups with and without (a) SGRE events, (b) DH type II bursts, and (c) major SEP events observed. {For SEP events we analyzed major events observed by GOES only and the second group of major SEP events observed by GOES or STEREO spacecraft.} Our statistical analysis show that all metric type II-associated HCMEs with an SGRE event have considerably higher initial acceleration and {also space speed if an major SEP event was also detected} than those of metric type II-associated HCMEs without an SGRE event. {The average space speeds of the SGRE-associated HCMEs without an SEP event and the non-SGRE-associated HCMEs with an major SEP event were similar.} The analysis of the HCMEs associated with only DH type II emission shows that the {three} SGRE-producing HCMEs without an SEP event {observed by GOES spacecraft} have higher space speed than any studied population of HCMEs not associated with an SGRE event. {However, one of those three HCMEs had an major SEP event observed by STEREO-B and the other two had elevated backgrounds at leat at the best connected spacecraft, so all three events could have accelerated protons.} Whereas the avergae initial acceleration is slightly lower than those of the metric type II-associated HCMEs without an SGRE event, but clearly higher than those of the DH type II-associated HCMEs without an SGRE event. The lower value is expected because CMEs associated with only a DH type II radio burst accelerate slowly and the shock forms later. This result resembles the kinematic hierarchy of CMEs with major SEP events, where rare, slowly accelerating but eventually fast CMEs associated with filament eruptions outside active regions can produce large SEP events at 1 au. In the case filament eruptions, we know that the resulting energy spectrum is soft, but the general idea that occasionally the initial acceleration of the CME is slower, but acceleration continues long enough so that a sufficiently strong shock forms later at higher altitudes is comparable. In general, our results are similar to those reported by \cite{2022CEAB..46...1G}. The SGRE-associated HCMEs seem to conform to the hierarchy between the initial acceleration and speed of the CME and the fluence spectral index as described by \cite{2016ApJ...833..216G}. Clearly, the existence of $>$300 MeV protons is connected to a high initial acceleration and speed of the associated HCME. Therefore, our results suggest that CME-driven shocks are the likely source for the $>$300 MeV protons required to produce SGREs at the Sun.

The mirror effect near the Sun limits the number of protons that can penetrate deep enough, i.e., particles with a pitch angle $\alpha$ in the sheath region cannot penetrate a near-Sun region if $\mu=\cos{\alpha}$ is larger than the critical value $\mu_{c}$:
\begin{equation}
  \left|\mu\right| \geq \mu_c \equiv \sqrt{1-\frac{B_{sheath}}{B_\odot}}.
\end{equation}

Because the foot points of the field lines crossing the shock nose could be connected to areas outside the source active region where the average magnetic field strength $B_\sun$ is considerably lower than in active regions, the mirror ratio $B_{sheath}/B_\sun$ increases and the width of the loss cone $\alpha_{c}=\cos^{-1}{\mu_{c}}=\sin^{-1}{\sqrt{B_{sheath}/B_\sun}}$ becomes larger. {Because the CME flux ropes have a pile-up region in front of them, the magnetic field within the sheath could be significantly larger than the ambient field, which will further lower the mirror ratio.} Therefore, more protons can precipitate deep into solar atmosphere. As mentioned earlier, enhanced turbulence increases scattering into the loss cone, which in turn increases the number of precipitating particles at the foot points. The level of the turbulence and its time evolution along the flux-rope-wrapping field lines and in the atmospheric layers close to the Sun are difficult to estimate. For example, EUV waves associated with large solar eruptions and propagating long distances over the solar surface clearly indicate that coronal shocks and CME lateral expansion affect the solar atmosphere far from the eruption site, most likely resulting in large volumes of enhanced turbulence around the source active region. 

It should be noted that fast CME shocks are the only sites for which we have clear corroborating observational evidence for acceleration of $>$300 MeV protons over extended times long after the end of the solar flare. \cite{2018ApJ...864...39W} studied the properties of the soft X-ray flares, CMEs and SEP events associated with SGRE events. They found that SGRE events are not produced by the brightest, most intense X-ray flares. In their reverse study, they found that during the period from 2011 March to 2015 June 45 X-class soft X-ray flares were detected, but only 15 of those were associated with a SGRE event. Similarly, their study showed that SGRE events are associated with fast CMEs and the SGRE duration increases as the CME speed increases. The reverse study of the fast HCMEs with speeds above 1500 km s$^{-1}$ found only four HCMEs without a reported gamma-ray event. Two of the four HCMEs, the 2011 September 22 10:48 UT and 2012 July 19 05:24 UT halos, had concurrent Fermi/LAT observations. In both events, the LAT spectra showed slight increases that were not significant enough to be characterized as detection. In the study of related SEP events observed by GOES, \cite{2018ApJ...864...39W} list only the 2011 March 07 SGRE event as a magnetically well-connected to GOES and without a significant background increase due to a preceding event but did not show any increase in the GOES $>$300 MeV flux. \cite{2019JPhCS1332a2004G} suggest that the lack of high-energy protons is due to a poor latitudinal magnetic connection of the shock nose to Earth because the flare occurred at the heliographic latitude of N31 and the northern polar region of the Sun is tilted away from Earth in March. Similarly, \cite{2021arXiv210811286G} showed that soft energy spectrum observed by GOES during the  2014 January 7 SGRE event was due to poor magnetic connectivity of the shock nose to an Earth observer. The final conclusion of \cite{2018ApJ...864...39W} is that their results favor the CME-shock as the source of the SGRE-producing protons.

The $>$300 MeV protons are accelerated near the nose of the CME-driven shock \citep[e.g.,][]{2013ApJ...765L..30G,2014EP&S...66..104G,2014ASPC..484...63G}{, with a possible exception of the earliest phase of the eruption, where the fast lateral expansion of the shock could result in efficient particle acceleration away from the nose region}. Some fraction of the shock-accelerated protons escapes into the IP space and are detected as an SEP event, but others propagate along the magnetic field lines deep down into the solar atmosphere and generate SGRE. In addition, the magnetic field lines that are pushed ahead of the CME body maintain a continuous connection between the shock and the solar atmosphere. Therefore, if the shock can accelerate protons to energies $>$300 MeV, protons will have a propagation path back to the Sun and can generate SGREs. The key aspect of the CME-shock model is that the magnetic field lines protons travel along sunwards cross the shock front into the sheath region behind the shock, wrap around the CME flux rope and connect back to the Sun in areas outside the foot points of the CME flux rope and possibly also areas outside the source active region. Therefore, the locations of the foot points are widely separated providing a natural explanation for the spatially extended source of gamma-ray emission.

We estimated the radial distance the CME-driven shock at the end time of the SGRE event for the 2012 January 23 and 2012 March 07 SGRE events. These events were associated with the two longest duration type II radio bursts. We estimated the shock radial distance by forward fitting a spheroidal shock model to STEREO/HI white-light images of the CME and obtained for the 2012 January 23 SGRE event the shock radial distance $r=121$ R$_\sun$ and for the 2012 March 07 SGRE event $r=140$ R$_\sun$. In addition, we used the frequency of the type II radio burst obtained from the radio dynamic spectra of Wind/WAVES together with a radial density model to get another estimate for the shock radial distance. The obtained distances from radio measurement were slightly longer, for the 2012 January 23 SGRE event $r=132$ R$_\sun$, and for the 2012 March 07 SGRE event $r=173$ R$_\sun$. 

\cite{2016ApJ...817...14P} estimated the speed of the 2012 March 07 CME using the standard aerodynamic drag-force model approach, where the CME travelled through quiet or perturbed solar wind (SW). Based on their results (their Figure 8), the estimated speed of the CME at 22 UT for the quiet SW model was $\approx$740 km s$^{-1}$ and for the perturbed SW  model $\approx$820 km s$^{-1}$. The perturbed SW model matches better the CME arrival time and speed at the Wind spacecraft. Therefore, our estimated CME speed of 750 km s$^{-1}$ seems to be slightly below the one obtained from the perturbed SW model. Recently, \cite{2023FrASS...949906S} studied the arrival signatures of the 2012 March 07 CME at several heliospheric locations. They report that Venus Express detected the arrival of the CME ejecta at 13:28 UT, when Venus was at the radial distance of 154 R$_\sun$. Therefore, the radial distance estimated from the forward fitting of the spheroidal shock model, r=140 R$_\sun$, is clearly too low and most likely the radial distance obtained from radio observations, r=173 R$_\sun$, is closer to the actual distance. The model fitting to images of the 2012 March 07 CME was difficult because the CME structure was very faint in the STEREO-B images (see Figure \ref{fig:shockfit}c). {Therefore, the sensitivity of the imaging system may not be high enough to detect the shock in front of the CME.}

The estimation of the radial distance of the shock at the end time of the SGRE event is quite complicated. The location and the shape of the shock front is difficult to discern from white-light images of the CME \citep[e.g.,][]{2009SoPh..259..297O,2019SpWea..17..539S}. CMEs are transparent structures and intensity of Thomson scattering depends on viewing angle relative to the structure. The location of type II radio source on the shock front is also difficult to measure. Imaging radio instrument operate at the higher frequencies, which correspond to heights of couple solar radius above the solar surface. Direction finding and triangulation can be used to locate the interplanetary type II radio sources at lower frequencies. However, scattering of radio waves and low intensity of type II emission limit the accuracy of the direction-finding measurements. 

\begin{figure}[ht!]
\plotone{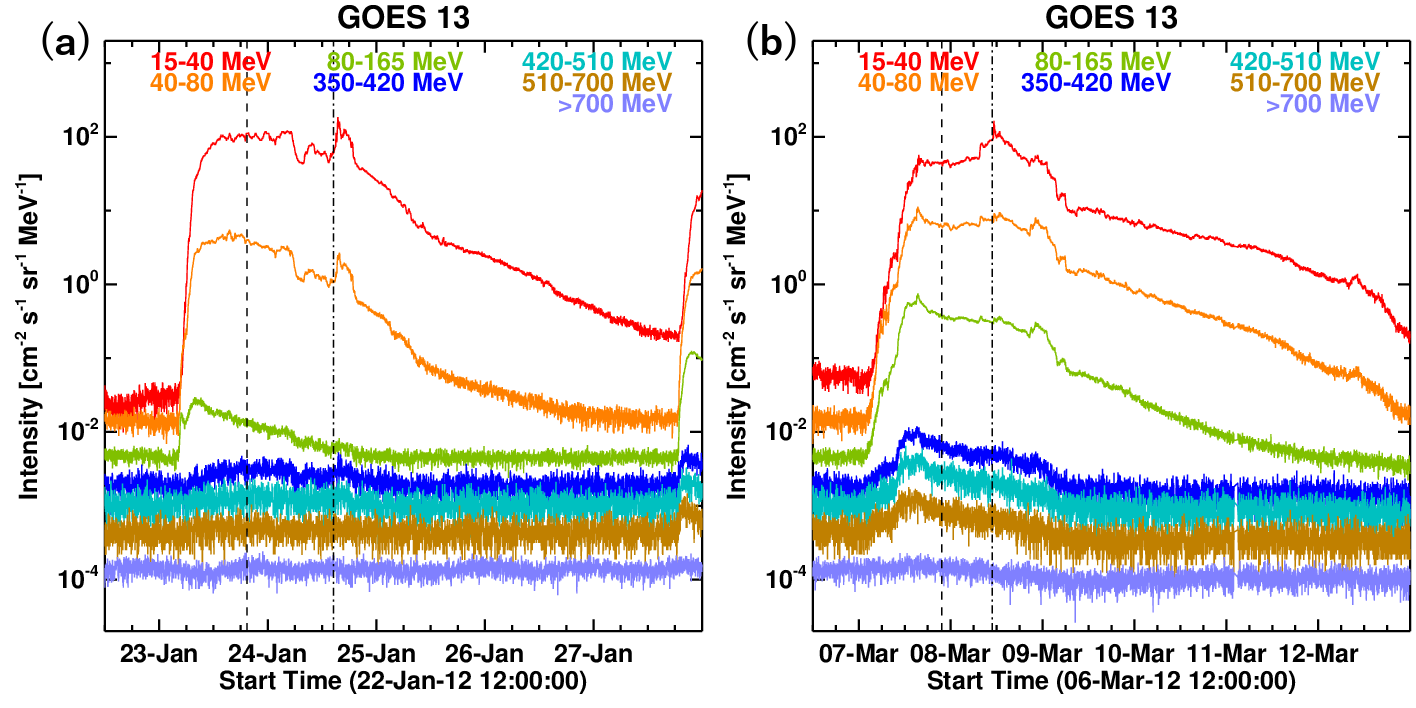}
\caption{GOES-13 measurements of the SEP events associated with the 2012 January 23 (a) and 2012 March 07 CMEs and SGRE events. Both events show enhanced intensity in the 350--420 MeV channel (blue). The March 07 event has a clear flux increase in the higher-energy 510--700 MeV channel (brown). The estimated SGRE end times at 2012 January 23 19:25 UT and March 07 21:40 UT are marked by dashed lines and the SOHO shock times at 2012 January 24 14:33 UT and March 08 10:53 UT by dash-dot lines.  \label{fig:goes}}
\end{figure}

The CME-driven shock in both events reached SOHO spacecraft (2012 January 24 14:33 UT and 2012 March 08 10:53 UT respectively) and when shocks passed 1 au about 30 minutes later, GOES spacecraft observed a clear increase around the shock time visible in Figure \ref{fig:goes}. Both events had particle flux increase in the GOES 350--429 MeV channels indicating that the CME-driven shock did accelerate $>$300 MeV protons. In both cases the 1-au enhancement continued beyond the end times of the SGRE events, estimated to be at 2012 January 23 19:25 UT and 2012 March 07 21:40 UT, respectively. During the March 07 event, the particle increase was detectable at even higher energies, up to the 510--700 MeV energy range. The event integrated fluence spectrum of 2012 January 23 event provided by PAMELA indicate that SEP flux at 1 au extended above 300 MeV \citep{2018ApJ...862...97B}. Therefore, the CME-driven shock clearly must accelerate $>$300 MeV protons far away from the Sun, providing support for the CME-shock as the source of the SGRE-producing protons.

The 2012 March 07 SGRE event was very bright that the location of the $>$100 MeV gamma-ray emission source could be estimated over several time intervals over a period of about 10 hours \citep{2014ApJ...789...20A,2021ApJS..252...13A}. The emission centroid seemed to move away from the flare site across the solar towards west. \cite{2021ApJS..252...13A} studied two other, bright SGRE flares on 2014 February 25 and 2017 September 10. The gamma-ray intensity of the 2014 February flare was weaker, so they could determine the location of the emission centroid only during two intervals over three hours. The September 2017 flare was brighter, and the location was determined in three intervals over 7 hours, but the flare occurred over the western limb of the Sun making the detection of possible source movement difficult. In both events the centroid remained consistent with the AR location. Therefore, the movement of the SGRE source during the 2012 March 07 event supports CME-shock scenario, whereas the detection of a possible source movement in the 2014 February 25 and 2017 September 10 events is complicated because of the weaker gamma-ray intensity or unfavorable location of the flare.

\section{Conclusions}

We compared acceleration and speed of CMEs associated with gamma-ray flares with 'Prompt' and/or 'Delayed' (SGRE event) component as defined by \cite{2021ApJS..252...13A}. In addition, we divided the on-disk HCMEs associated with type II radio bursts into groups with or without SGRE events, SEP events, metric or DH type II radio bursts and compared the average acceleration and speed between the HCME groups. We showed that the CMEs associated with the 'Delayed' gamma-ray component and the metric type II-producing HCMEs associated with SGRE events together with a DH type II radio burst and/or a major SEP events have higher initial acceleration and space speed than the CMEs associated with the 'Prompt-only' gamma-ray component or the SEP- or type II-associated HCMEs without SGRE. {The only exception was the space speed of metric type II -associated HCMEs with an major SEP event but without an SGRE event that had similar average space speeds as the SGRE-associated HCMEs without an major SEP event.} Similar high initial acceleration and fast speed characteristics are shared by CMEs associated with GLEs, which are guaranteed to have $>$300 MeV protons. The SGRE-associated CMEs also conform to the hierarchy between the initial acceleration and speed of the CME and the fluence spectral index as described by \cite{2016ApJ...833..216G}. Therefore, our findings support the CME-driven shock as the source of $>$300 MeV protons producing SGRE events.

We estimated the radial distance the CME-driven shock at the end time of the SGRE event with the long-duration type II radio bursts on 2012 January 23 and 2012 March 07 using STEREO/HI white-light images of the CME and radio dynamic spectra of Wind/WAVES. The shock radial distances for 2012 January 23 SGRE event were r=121 R$_\sun$ and r=132 R$_\sun$, and for the 2012 March 07 SGRE event r=140 R$_\sun$ and r=173 R$_\sun$, respectively. The distances derived from white-light and radio observations are reasonably consistent, indicating that the radio source is near the shock nose as assumed. The distances are also consistently longer that the estimated shock height $\approx$70 R$_\sun$ for the shorter-duration 2014 February 25 SGRE event \cite{2019JPhCS1332a2004G} (Gopalswamy et al. 2019). Because the shock location is not visible in the white-light images, the radial distance estimated from forward fitting of the spheroidal shock model are probably underestimations. At the end time of the SGRE event, the shock speeds were still high enough (975 km s$^{-1}$ and 750 km s$^{-1}$) for high-energy particle acceleration. Therefore, we conclude that strong CME-driven shocks accelerate $>$300 MeV protons up to the radial distances of 0.6--0.8 au. 

\begin{acknowledgments}
We thank the Fermi/LAT, GOES, SOHO/LASCO, STEREO/SECCHI, Wind/WAVES, and HELCATS teams for providing the data. PM and SA were partially supported by NSF grant AGS-2043131. NG was supported by NASA's STEREO project and the Living With a Star program. HX was partially supported by NSF grant AGS-2228967.
\end{acknowledgments}

\appendix

\section{CMEs and X-ray Flares Associated with On-disk Fermi/LAT Solar Flares \label{sec:A}}

Table \ref{tab:tbl1_data} contains data for the CME and X-ray flare data used in Table \ref{tab:tbl2}. The first column gives the first observation date and time of the CME followed by the measured sky-plane speed and the projection-corrected space speed in the second and third columns. The fourth column lists the estimated initial acceleration of the CME. The columns 5--7 list the location in heliographic coordinates, the onset and peak times of the GOES soft X-ray flare. The last column lists the gamma-ray components detected by Fermi/LAT taken from \cite{2021ApJS..252...13A}. 

\begin{deluxetable*}{lcccrlll}
\tablecaption{CME and X-ray flare data for Fermi/LAT solar flares \label{tab:tbl1_data}}
\tablewidth{700pt}
\tabletypesize{\scriptsize}
\tablenum{A}
\tablehead{
\colhead{CME} & \colhead{Sky Speed} &
\colhead{Space Speed} & \colhead{Acc} &
\colhead{Location} & \colhead{Flare Onset} &
\colhead{Flare Peak} & \colhead{Gamma-ray Components} \\ 
\colhead{(UT)} & \colhead{(km $^{-1}$)} &
\colhead{(km s$^{-1}$)} & \colhead{(km s$^{-2}$)} & 
\colhead{(UT)} & \colhead{(UT)} &
\colhead{(UT)} & \colhead{}
}
\colnumbers
\startdata
2010/06/12 01:31 &  620 &  674 &  0.42 &  N23W43 & 2010/06/12 00:30 & 2010/06/12 00:57 &                LLE-Prompt \\
2011/03/07 20:00 & 2125 & 2223 &  1.28 &  N30W48 & 2011/03/07 19:43 & 2011/03/07 20:12 &                   Delayed \\
2011/06/07 06:49 & 1255 & 1321 &  0.88 &  S21W54 & 2011/06/07 06:16 & 2011/06/07 06:41 &                   Delayed \\
2011/08/04 04:12 & 1315 & 1477 &  1.54 &  N19W36 & 2011/08/04 03:41 & 2011/08/04 03:57 &                   Delayed \\
2011/08/09 08:12 & 1610 & 1640 &  1.61 &  N17W69 & 2011/08/09 07:48 & 2011/08/09 08:05 &      Prompt Short-Delayed \\
2011/09/06 23:05 &  575 &  830 &  1.73 &  N14W18 & 2011/09/06 22:12 & 2011/09/06 22:20 &  LLE-Prompt Short-Delayed \\
2011/09/07 23:05 &  710 &  735 &  2.04 &  N14W28 & 2011/09/07 22:32 & 2011/09/07 22:38 &                   Delayed \\
2011/09/24 09:48 & 1936 & 2235 &  1.96 &  N12E60 & 2011/09/24 09:21 & 2011/09/24 09:40 &  LLE-Prompt Short-Delayed \\
2012/01/23 04:00 & 2175 & 2511 &  1.99 &  N28W21 & 2012/01/23 03:38 & 2012/01/23 03:59 &                   Delayed \\
2012/01/27 18:27 & 2508 & 2541 &  0.71 &  N27W78 & 2012/01/27 17:37 & 2012/01/27 18:37 &                   Delayed \\
2012/03/05 04:00 & 1531 & 1627 &  0.52 &  N17E52 & 2012/03/05 03:17 & 2012/03/05 04:09 &                   Delayed \\
2012/03/07 00:24 & 2684 & 3146 &  2.38 &  N17E27 & 2012/03/07 00:02 & 2012/03/07 00:24 &                   Delayed \\
2012/03/09 04:26 &  950 & 1229 &  0.66 &  N15W03 & 2012/03/09 03:22 & 2012/03/09 03:53 &         No-Prompt Delayed \\
2012/03/10 18:00 & 1296 & 1638 &  0.94 &  N17W24 & 2012/03/10 17:15 & 2012/03/10 17:44 &                   Delayed \\
2012/05/17 01:48 & 1582 & 1596 &  1.21 &  N11W76 & 2012/05/17 01:25 & 2012/05/17 01:47 &                   Delayed \\
2012/06/03 18:12 &  772 &  786 &  1.87 &  N16E38 & 2012/06/03 17:48 & 2012/06/03 17:55 &  LLE-Prompt Short-Delayed \\
2012/07/06 23:24 & 1828 & 1907 &  4.54 &  S13W59 & 2012/07/06 23:01 & 2012/07/06 23:08 &                   Delayed \\
2012/08/06 05:12 &  198 &  199 &  0.66 &  S14E84 & 2012/08/06 04:33 & 2012/08/06 04:38 &                LLE-Prompt \\
2012/11/13 02:24 &  980 & 1002 &  2.78 &  S25E46 & 2012/11/13 01:58 & 2012/11/13 02:04 &                    Prompt \\
2013/04/11 07:24 &  861 & 1369 &  1.09 &  N09E12 & 2013/04/11 06:55 & 2013/04/11 07:16 &   No-Prompt Short-Delayed \\
2013/05/13 02:00 & 1270 & 1270 &  0.88 &  N11E90 & 2013/05/13 01:53 & 2013/05/13 02:17 &                   Delayed \\
2013/05/13 16:07 & 1850 & 1852 &  1.82 &  N11E85 & 2013/05/13 15:48 & 2013/05/13 16:05 &                   Delayed \\
2013/05/14 01:25 & 2625 & 2645 &  4.01 &  N08E77 & 2013/05/14 01:00 & 2013/05/14 01:11 &         No-Prompt Delayed \\
2013/05/15 01:48 & 1366 & 1408 &  0.71 &  N12E64 & 2013/05/15 01:15 & 2013/05/15 01:48 &         No-Prompt Delayed \\
2013/10/25 08:12 &  587 &  599 &  1.25 &  S08E73 & 2013/10/25 07:53 & 2013/10/25 08:01 &                   Delayed \\
2013/10/28 02:24 &  695 &  726 &  0.55 &  N04W66 & 2013/10/28 01:41 & 2013/10/28 02:03 &                LLE-Prompt \\
2013/10/28 04:48 & 1201 & 1270 &  2.35 &  N08W71 & 2013/10/28 04:32 & 2013/10/28 04:41 &                LLE-Prompt \\
2013/10/28 15:36 &  812 & 1098 &  2.29 &  S06E28 & 2013/10/28 15:07 & 2013/10/28 15:15 &                   Delayed \\
2013/10/28 21:25 &  771 &  777 &  1.44 &  N07W83 & 2013/10/28 20:48 & 2013/10/28 20:57 &                LLE-Prompt \\
2014/01/07 18:24 & 1830 & 2246 &  1.34 &  S15W11 & 2014/01/07 18:04 & 2014/01/07 18:32 &                   Delayed \\
2014/02/25 01:25 & 2147 & 2153 &  3.59 &  S12E82 & 2014/02/25 00:39 & 2014/02/25 00:49 &        LLE-Prompt Delayed \\
2014/06/10 13:30 & 1469 & 1473 &  1.53 &  S17E82 & 2014/06/10 12:36 & 2014/06/10 12:52 &        LLE-Prompt Delayed \\
2014/06/11 09:24 &  829 &  915 &  2.18 &  S18E65 & 2014/06/11 08:59 & 2014/06/11 09:06 &             Short-Delayed \\
2014/09/10 18:00 & 1267 & 1652 &  1.15 &  N14E02 & 2014/09/10 17:21 & 2014/09/10 17:45 &             Short-Delayed \\
2015/06/21 02:36 & 1366 & 1740 &  0.97 &  N12E16 & 2015/06/21 02:06 & 2015/06/21 02:36 &            Prompt Delayed \\
2015/06/25 08:36 & 1627 & 1805 &  2.15 &  N09W42 & 2015/06/25 08:02 & 2015/06/25 08:16 &                   Delayed \\
2017/09/06 12:24 & 1571 & 1819 &  3.37 &  S08W33 & 2017/09/06 11:53 & 2017/09/06 12:02 &                   Delayed \\
2017/09/10 16:00 & 3163 & 3163 &  1.70 &  S09W92 & 2017/09/10 15:35 & 2017/09/10 16:06 &            Prompt Delayed \\
\enddata
\tablecomments{Gamma-ray Components are taken from \cite{2021ApJS..252...13A}.}
\end{deluxetable*}

\section{Cycle 24 HCMEs with Type II Radio Bursts and on-disk X-ray Flares \label{sec:B}}

Table \ref{tab:tbl2_data} contains data for the cycle-24 HCME and X-ray flares used in Tables \ref{tab:tbl2} and \ref{tab:tbl3}. The columns 1--7 are the same as in Table \ref{tab:tbl1_data}. Columns 8--9 list the onset times of the reported metric and DH type II radio bursts. {The DH type II onset times are listed for Wind/WAVES, except on 03 August 2011, when only STEREO-A/WAVES detected a DH type II burst. The column 10 indicates on which spacecraft the WAVES instruments could detect a DH type II radio burst (W=Wind, A=STEREO-A, B=STEREO-B, '-'=no report). The columns 11--12 mark if the event had a major SEP event (G=GOES, A=STEREO-A, B=STEREO-B, '-'=data gap)} and an SGRE event ('Delayed' component detected), respectively.

\begin{longrotatetable}
\begin{deluxetable*}{lcccrllllccc}
\tablecaption{Cycle 24 HCMEs with type II radio bursts \label{tab:tbl2_data}}
\tablewidth{700pt}
\tabletypesize{\scriptsize}
\tablenum{B}
\tablehead{
\colhead{HCME} & \colhead{Sky Speed} &
\colhead{Space Speed} & \colhead{Acc} &
\colhead{Location} & \colhead{Flare Onset} &
\colhead{Flare Peak} & \colhead{m-Type II} &
\colhead{DH-Type II} &\colhead{WAVES S/C} & \colhead{SEP} & \colhead{SGRE} \\ 
\colhead{(UT)} & \colhead{(km $s^{-1}$)} &
\colhead{(km s$^{-1}$)} & \colhead{(km s$^{-2}$)} & 
\colhead{} & \colhead{(UT)} &
\colhead{(UT)} & \colhead{(UT)} &
\colhead{(UT)} & \colhead{} & \colhead{G/A/B} & \colhead{}
}
\colnumbers
\startdata
2010/08/01 13:42 &  850 & 1030 &  0.34 & N20E36 & 2010/08/01 07:36 & 2010/08/01 08:26 &          \nodata & 2010/08/01 09:20 & W/A/B & 0/-/0 & 0 \\
2010/08/07 18:36 &  871 & 1102 &  0.63 & N11E34 & 2010/08/07 17:55 & 2010/08/07 18:24 & 2010/08/07 18:08 & 2010/08/07 18:35 & W/A/B & 0/0/1 & 0 \\
2010/08/14 10:12 & 1205 & 1280 &  0.51 & N17W52 & 2010/08/14 09:23 & 2010/08/14 10:05 & 2010/08/14 09:52 &          \nodata & -/-/- & 1/0/0 & 0 \\
2011/02/14 18:24 &  326 &  544 &  1.51 & S20W04 & 2011/02/14 17:20 & 2011/02/14 17:26 & 2011/02/14 17:28 &          \nodata & -/-/- & 0/0/0 & 0 \\
2011/02/15 02:24 &  669 &  960 &  1.33 & S20W10 & 2011/02/15 01:44 & 2011/02/15 01:56 & 2011/02/15 01:52 & 2011/02/15 02:10 & W/A/B & 0/0/1 & 0 \\
2011/03/07 20:00 & 2125 & 2223 &  1.28 & N30W48 & 2011/03/07 19:43 & 2011/03/07 20:12 & 2011/03/07 19:54 & 2011/03/07 20:00 & W/A/- & 1/1/0 & 1 \\
2011/06/02 08:12 &  976 & 1147 &  0.64 & S19E25 & 2011/06/02 07:16 & 2011/06/02 07:46 &          \nodata & 2011/06/02 08:00 & W/A/B & 0/0/0 & 0 \\
2011/06/07 06:49 & 1255 & 1321 &  0.88 & S21W54 & 2011/06/07 06:16 & 2011/06/07 06:41 & 2011/06/07 06:25 & 2011/06/07 06:45 & W/A/B & 1/0/0 & 1 \\
2011/06/21 03:16 &  719 &  882 &  0.12 & N16W08 & 2011/06/21 01:22 & 2011/06/21 03:25 &          \nodata & 2011/06/21 03:07 & W/-/- & 0/0/0 & 0 \\
2011/08/03 14:00 &  610 &  785 &  0.37 & N16W30 & 2011/08/03 13:13 & 2011/08/03 13:48 & 2011/08/03 13:35 & 2011/08/03 13:38 & -/A/- & 0/0/0 & 0 \\
2011/08/04 04:12 & 1315 & 1477 &  1.54 & N19W36 & 2011/08/04 03:41 & 2011/08/04 03:57 & 2011/08/04 03:54 & 2011/08/04 04:15 & W/A/B & 1/0/0 & 1 \\
2011/08/09 08:12 & 1610 & 1640 &  1.61 & N17W69 & 2011/08/09 07:48 & 2011/08/09 08:05 & 2011/08/09 08:01 & 2011/08/09 08:20 & W/-/- & 1/0/0 & 1 \\
2011/09/06 02:24 &  782 & 1232 &  1.37 & N14W07 & 2011/09/06 01:35 & 2011/09/06 01:50 & 2011/09/06 01:46 & 2011/09/06 02:00 & W/-/- & 0/0/0 & 0 \\
2011/09/06 23:05 &  575 &  830 &  1.73 & N14W18 & 2011/09/06 22:12 & 2011/09/06 22:20 & 2011/09/06 22:19 & 2011/09/06 22:30 & W/A/- & 0/0/0 & 1 \\
2011/09/22 10:48 & 1905 & 1905 &  0.99 & N09E89 & 2011/09/22 10:29 & 2011/09/22 11:01 & 2011/09/22 10:39 & 2011/09/22 11:05 & W/-/B & 1/1/1 & 0 \\
2011/09/24 12:48 & 1915 & 2018 &  0.58 & N10E56 & 2011/09/24 12:22 & 2011/09/24 13:20 &          \nodata & 2011/09/24 12:50 & W/-/B & 0/0/0 & 0 \\
2011/09/24 19:36 &  972 & 1076 &  1.49 & N12E42 & 2011/09/24 19:09 & 2011/09/24 19:21 & 2011/09/24 19:14 &          \nodata & -/-/- & 0/0/1 & 0 \\
2011/11/09 13:36 &  907 & 1012 &  0.54 & N24E35 & 2011/11/09 13:04 & 2011/11/09 13:35 & 2011/11/09 13:11 & 2011/11/09 13:30 & W/-/B & 0/0/0 & 0 \\
2011/11/26 07:12 &  933 & 1001 &  0.60 & N17W49 & 2011/11/26 06:42 & 2011/11/26 07:10 &          \nodata & 2011/11/26 07:15 & W/A/- & 1/1/0 & 0 \\
2012/01/19 14:36 & 1120 & 1269 &  0.15 & N32E22 & 2012/01/19 13:44 & 2012/01/19 16:05 &          \nodata & 2012/01/19 15:00 & W/A/B & 0/0/1 & 0 \\
2012/01/23 04:00 & 2175 & 2511 &  1.99 & N28W21 & 2012/01/23 03:38 & 2012/01/23 03:59 & 2012/01/23 03:43 & 2012/01/23 04:00 & W/A/- & 1/1/1 & 1 \\
2012/01/27 18:27 & 2508 & 2541 &  0.71 & N27W78 & 2012/01/27 17:37 & 2012/01/27 18:37 & 2012/01/27 18:10 & 2012/01/27 18:30 & W/A/B & 1/1/0 & 1 \\
2012/03/05 04:00 & 1531 & 1627 &  0.52 & N17E52 & 2012/03/05 03:17 & 2012/03/05 04:09 &          \nodata & 2012/03/05 04:00 & W/A/B & 0/0/0 & 1 \\
2012/03/07 00:24 & 2684 & 3146 &  2.38 & N17E27 & 2012/03/07 00:02 & 2012/03/07 00:24 & 2012/03/07 00:17 & 2012/03/07 01:00 & W/A/B & 1/0/1 & 1 \\
2012/03/07 01:30 & 1825 & 2160 &  4.00 & N15E26 & 2012/03/07 01:05 & 2012/03/07 01:14 & 2012/03/07 01:09 &          \nodata & -/-/- & 0/0/0 & 0 \\
2012/03/09 04:26 &  950 & 1229 &  0.66 & N15W03 & 2012/03/09 03:22 & 2012/03/09 03:53 & 2012/03/09 03:43 & 2012/03/09 04:10 & W/-/- & 0/1/0 & 1 \\
2012/03/10 18:00 & 1296 & 1638 &  0.94 & N17W24 & 2012/03/10 17:15 & 2012/03/10 17:44 &          \nodata & 2012/03/10 17:55 & W/A/- & 0/0/0 & 1 \\
2012/03/13 17:36 & 1884 & 1931 &  0.89 & N17W66 & 2012/03/13 17:05 & 2012/03/13 17:41 & 2012/03/13 17:15 & 2012/03/13 17:35 & W/A/- & 1/0/0 & 0 \\
2012/04/05 21:25 &  828 & 1065 &  0.66 & N18W29 & 2012/04/05 20:43 & 2012/04/05 21:10 & 2012/04/05 21:08 &          \nodata & -/-/- & 0/0/0 & 0 \\
2012/04/09 12:36 &  921 &  945 &  0.38 & N20W65 & 2012/04/09 12:02 & 2012/04/09 12:44 & 2012/04/09 12:28 & 2012/04/09 12:20 & W/A/- & 0/0/0 & 0 \\
2012/04/23 18:24 &  528 &  769 &  0.99 & N14W17 & 2012/04/23 17:38 & 2012/04/23 17:51 & 2012/04/23 17:42 &          \nodata & -/-/- & 0/0/0 & 0 \\
2012/05/17 01:48 & 1582 & 1596 &  1.21 & N11W76 & 2012/05/17 01:25 & 2012/05/17 01:47 & 2012/05/17 01:31 & 2012/05/17 01:40 & W/A/- & 1/0/0 & 1 \\
2012/07/04 17:24 &  662 &  830 &  2.31 & N14W34 & 2012/07/04 16:33 & 2012/07/04 16:39 & 2012/07/04 16:42 & 2012/07/04 17:00 & W/-/- & 0/0/0 & 0 \\
2012/07/06 23:24 & 1828 & 1907 &  4.54 & S13W59 & 2012/07/06 23:01 & 2012/07/06 23:08 & 2012/07/06 23:09 & 2012/07/06 23:10 & W/A/- & 1/0/0 & 1 \\
2012/07/12 16:48 &  885 & 1405 &  0.51 & S15W01 & 2012/07/12 16:03 & 2012/07/12 16:49 & 2012/07/12 16:25 & 2012/07/12 16:45 & W/-/- & 1/0/1 & 0 \\
2012/07/19 05:24 & 1631 & 1631 &  0.37 & S13W88 & 2012/07/19 04:45 & 2012/07/19 05:58 & 2012/07/19 05:24 & 2012/07/19 05:30 & W/-/- & 1/0/0 & 0 \\
2012/07/28 21:12 &  420 &  463 &  0.64 & S25E54 & 2012/07/28 20:44 & 2012/07/28 20:56 & 2012/07/28 20:52 &          \nodata & -/-/- & 0/0/0 & 0 \\
2012/07/31 11:24 &  567 &  605 &  0.23 & N19E59 & 2012/07/31 10:46 & 2012/07/31 11:30 & 2012/07/31 11:04 &          \nodata & -/-/- & 0/0/0 & 0 \\
2012/08/13 13:25 &  435 &  705 &  1.68 & N22W03 & 2012/08/13 12:33 & 2012/08/13 12:40 & 2012/08/13 12:41 &          \nodata & -/-/- & 0/0/0 & 0 \\
2012/08/31 20:00 & 1442 & 1495 &  0.35 & S19E50 & 2012/08/31 19:32 & 2012/08/31 20:43 & 2012/08/31 19:42 & 2012/08/31 20:00 & W/A/- & 1/0/1 & 0 \\
2012/09/28 00:12 &  947 & 1093 &  0.87 & N09W31 & 2012/09/27 23:36 & 2012/09/27 23:57 & 2012/09/27 23:44 & 2012/09/27 23:55 & W/A/- & 1/0/1 & 0 \\
2012/11/08 02:36 &  855 &  855 &  0.95 & N13E89 & 2012/11/08 02:08 & 2012/11/08 02:23 & 2012/11/08 02:21 &          \nodata & -/-/- & 0/0/0 & 0 \\
2012/11/21 16:00 &  529 &  942 &  0.79 & N05E05 & 2012/11/21 15:10 & 2012/11/21 15:30 & 2012/11/21 15:33 &          \nodata & -/-/- & 0/0/0 & 0 \\
2013/03/15 07:12 & 1063 & 1366 &  0.39 & N11E12 & 2013/03/15 06:00 & 2013/03/15 06:58 &          \nodata & 2013/03/15 07:00 & W/-/- & 1/0/0 & 0 \\
2013/04/11 07:24 &  861 & 1369 &  1.09 & N09E12 & 2013/04/11 06:55 & 2013/04/11 07:16 & 2013/04/11 07:02 & 2013/04/11 07:10 & W/-/B & 1/0/1 & 1 \\
2013/05/13 02:00 & 1270 & 1270 &  0.88 & N11E90 & 2013/05/13 01:53 & 2013/05/13 02:17 & 2013/05/13 02:10 & 2013/05/13 02:20 & W/-/B & 0/0/1 & 1 \\
2013/05/13 16:07 & 1850 & 1852 &  1.82 & N11E85 & 2013/05/13 15:48 & 2013/05/13 16:05 & 2013/05/13 15:57 & 2013/05/13 16:15 & W/A/B & 0/0/1 & 1 \\
2013/05/14 01:25 & 2625 & 2645 &  4.01 & N08E77 & 2013/05/14 01:00 & 2013/05/14 01:11 & 2013/05/14 01:07 & 2013/05/14 01:16 & W/A/B & 0/0/0 & 1 \\
2013/05/15 01:48 & 1366 & 1408 &  0.71 & N12E64 & 2013/05/15 01:15 & 2013/05/15 01:48 & 2013/05/15 01:37 & 2013/05/15 01:49 & W/-/- & 1/0/0 & 1 \\
2013/05/17 09:12 & 1345 & 1412 &  1.68 & N12E57 & 2013/05/17 08:43 & 2013/05/17 08:57 & 2013/05/17 08:50 &          \nodata & -/-/- & 0/0/0 & 0 \\
2013/05/22 13:25 & 1466 & 1491 &  0.71 & N15W70 & 2013/05/22 12:57 & 2013/05/22 13:32 & 2013/05/22 12:59 & 2013/05/22 13:10 & W/A/B & 1/1/0 & 0 \\
2013/06/28 02:00 & 1037 & 1254 &  0.91 & S18W19 & 2013/06/28 01:36 & 2013/06/28 01:59 &          \nodata & 2013/06/28 01:53 & W/-/- & 0/0/0 & 0 \\
2013/08/17 19:12 & 1202 & 1418 &  0.54 & S05W30 & 2013/08/17 18:49 & 2013/08/17 19:33 & 2013/08/17 18:56 & 2013/08/17 20:25 & W/-/- & 0/0/0 & 0 \\
2013/08/30 02:48 &  949 & 1031 &  0.31 & N15E46 & 2013/08/30 01:51 & 2013/08/30 02:46 & 2013/08/30 02:12 & 2013/08/30 02:34 & W/-/- & 0/0/0 & 0 \\
2013/09/29 22:12 & 1179 & 1370 &  0.21 & N17W29 & 2013/09/29 21:43 & 2013/09/29 23:31 & 2013/09/29 21:53 & 2013/09/29 21:53 & W/A/B & 1/0/0 & 0 \\
2013/10/22 21:48 &  459 & 1070 &  3.57 & N04W01 & 2013/10/22 21:15 & 2013/10/22 21:20 & 2013/10/22 21:21 & 2013/10/22 21:33 & W/-/- & 0/0/0 & 0 \\
2013/10/24 01:25 &  399 &  766 &  1.42 & S10E08 & 2013/10/24 00:21 & 2013/10/24 00:30 & 2013/10/24 00:31 &          \nodata & -/-/- & 0/0/0 & 0 \\
2013/10/25 08:12 &  587 &  599 &  1.25 & S08E73 & 2013/10/25 07:53 & 2013/10/25 08:01 & 2013/10/25 07:59 &          \nodata & -/-/- & 0/0/1 & 1 \\
2013/10/25 15:12 & 1081 & 1103 &  1.53 & S06E69 & 2013/10/25 14:51 & 2013/10/25 15:03 & 2013/10/25 14:58 & 2013/10/25 15:08 & W/-/B & 0/0/0 & 0 \\
2013/10/28 02:24 &  695 &  726 &  0.55 & N04W66 & 2013/10/28 01:41 & 2013/10/28 02:03 & 2013/10/28 02:00 &          \nodata & -/-/- & 0/0/0 & 0 \\
2013/10/28 15:36 &  812 & 1098 &  2.29 & S06E28 & 2013/10/28 15:07 & 2013/10/28 15:15 & 2013/10/28 15:10 & 2013/10/28 15:24 & W/-/- & 0/0/0 & 1 \\
2013/10/29 22:00 & 1001 & 1001 &  1.39 & N05W89 & 2013/10/29 21:42 & 2013/10/29 21:54 & 2013/10/29 21:48 &          \nodata & -/-/- & 0/0/0 & 0 \\
2013/11/19 10:36 &  740 &  761 &  1.06 & S14W70 & 2013/11/19 10:14 & 2013/11/19 10:26 & 2013/11/19 10:24 & 2013/11/19 10:39 & W/-/- & 0/0/0 & 0 \\
2013/12/07 07:36 & 1085 & 1165 &  1.62 & S16W49 & 2013/12/07 07:17 & 2013/12/07 07:29 & 2013/12/07 07:27 & 2013/12/07 07:43 & W/-/- & 0/0/0 & 0 \\
2014/01/07 18:24 & 1830 & 2246 &  1.34 & S15W11 & 2014/01/07 18:04 & 2014/01/07 18:32 & 2014/01/07 18:17 & 2014/01/07 18:33 & W/A/B & 1/1/1 & 1 \\
2014/01/20 22:00 &  721 &  750 &  0.18 & S07E67 & 2014/01/20 21:39 & 2014/01/20 22:49 &          \nodata & 2014/01/20 22:24 & W/-/- & 0/0/0 & 0 \\
2014/02/20 08:00 &  948 &  960 &  0.53 & S15W73 & 2014/02/20 07:26 & 2014/02/20 07:56 & 2014/02/20 07:45 & 2014/02/20 08:05 & W/-/- & 1/0/0 & 0 \\
2014/02/25 01:25 & 2147 & 2153 &  3.59 & S12E82 & 2014/02/25 00:39 & 2014/02/25 00:49 & 2014/02/25 00:56 & 2014/02/25 00:56 & W/A/B & 1/1/1 & 1 \\
2014/03/20 04:36 &  740 &  921 &  1.10 & S14E35 & 2014/03/20 03:42 & 2014/03/20 03:56 & 2014/03/20 03:52 &          \nodata & -/-/- & 0/0/0 & 0 \\
2014/03/29 18:12 &  528 &  679 &  0.87 & N11W32 & 2014/03/29 17:35 & 2014/03/29 17:48 & 2014/03/29 17:53 & 2014/03/29 17:59 & W/-/- & 0/0/0 & 0 \\
2014/04/02 13:36 & 1471 & 1564 &  0.55 & N11E53 & 2014/04/02 13:18 & 2014/04/02 14:05 & 2014/04/02 13:23 & 2014/04/02 13:42 & W/-/B & 0/0/1 & 0 \\
2014/04/18 13:25 & 1203 & 1359 &  0.71 & S20W34 & 2014/04/18 12:31 & 2014/04/18 13:03 & 2014/04/18 12:55 & 2014/04/18 13:05 & W/-/- & 1/0/0 & 0 \\
2014/06/10 13:30 & 1469 & 1473 &  1.53 & S17E82 & 2014/06/10 12:36 & 2014/06/10 12:52 &          \nodata & 2014/06/10 12:58 & W/-/B & 0/0/1 & 1 \\
2014/07/08 16:36 &  773 &  841 &  1.00 & N12E56 & 2014/07/08 16:06 & 2014/07/08 16:20 & 2014/07/08 16:14 &          \nodata & -/-/- & 0/-/0 & 0 \\
2014/08/01 18:36 &  789 & 1256 &  1.16 & S10E11 & 2014/08/01 17:55 & 2014/08/01 18:13 & 2014/08/01 18:18 & 2014/08/01 18:58 & W/-/- & 0/0/0 & 0 \\
2014/08/22 11:12 &  600 &  993 &  1.18 & N12E01 & 2014/08/22 10:13 & 2014/08/22 10:27 &          \nodata & 2014/08/22 10:37 & W/-/- & 0/-/0 & 0 \\
2014/08/24 12:36 &  551 &  569 &  0.56 & S07E75 & 2014/08/24 12:00 & 2014/08/24 12:17 & 2014/08/24 12:14 &          \nodata & -/-/- & 0/-/0 & 0 \\
2014/08/25 15:36 &  555 &  697 &  0.46 & N05W36 & 2014/08/25 14:46 & 2014/08/25 15:11 & 2014/08/25 15:08 & 2014/08/25 15:20 & W/-/- & 0/-/0 & 0 \\
2014/09/09 00:06 &  920 & 1080 &  0.33 & N12E29 & 2014/09/08 23:34 & 2014/09/09 00:29 &          \nodata & 2014/09/09 00:05 & W/-/- & 0/0/0 & 0 \\
2014/09/10 18:00 & 1267 & 1652 &  1.15 & N14E02 & 2014/09/10 17:21 & 2014/09/10 17:45 &          \nodata & 2014/09/10 17:45 & W/-/- & 1/-/0 & 1 \\
2014/12/17 05:00 &  587 &  855 &  0.46 & S20E09 & 2014/12/17 04:20 & 2014/12/17 04:51 & 2014/12/17 04:44 & 2014/12/17 05:00 & W/-/- & 0/-/- & 0 \\
2014/12/19 01:04 & 1195 & 1513 &  1.48 & S11E15 & 2014/12/18 21:41 & 2014/12/18 21:58 & 2014/12/18 22:22 & 2014/12/18 22:31 & W/-/- & 0/-/- & 0 \\
2014/12/21 12:12 &  669 &  906 &  0.28 & S14W25 & 2014/12/21 11:24 & 2014/12/21 12:17 &          \nodata & 2014/12/21 12:05 & W/-/- & 0/-/- & 0 \\
2015/02/09 23:24 & 1106 & 1148 &  0.53 & N12E61 & 2015/02/09 22:59 & 2015/02/09 23:35 & 2015/02/09 23:14 &          \nodata & -/-/- & 0/-/- & 0 \\
2015/03/07 22:12 & 1261 & 1304 &  0.59 & S19E74 & 2015/03/07 21:45 & 2015/03/07 22:22 & 2015/03/07 21:57 &          \nodata & -/-/- & 0/-/- & 0 \\
2015/03/10 00:00 &  995 & 1081 &  0.75 & S18E45 & 2015/03/09 23:29 & 2015/03/09 23:53 & 2015/03/10 00:05 & 2015/03/10 00:10 & W/-/- & 0/-/- & 0 \\
2015/03/10 03:36 & 1040 & 1156 &  3.85 & S15E40 & 2015/03/10 03:19 & 2015/03/10 03:24 & 2015/03/10 03:28 &          \nodata & -/-/- & 0/-/- & 0 \\
2015/03/15 01:48 &  719 &  932 &  0.27 & S22W25 & 2015/03/15 01:15 & 2015/03/15 02:13 & 2015/03/15 01:27 &          \nodata & -/-/- & 0/-/- & 0 \\
2015/04/23 09:36 &  857 &  864 &  0.29 & N12W89 & 2015/04/23 09:18 & 2015/04/23 10:07 & 2015/04/23 09:22 &          \nodata & -/-/- & 0/-/- & 0 \\
2015/05/05 22:24 &  715 &  721 &  2.00 & N15E79 & 2015/05/05 22:05 & 2015/05/05 22:11 & 2015/05/05 22:12 & 2015/05/05 22:24 & W/-/- & 0/-/- & 0 \\
2015/05/13 18:48 &  438 &  730 &  1.35 & N13W16 & 2015/05/13 18:09 & 2015/05/13 18:18 & 2015/05/13 18:21 &          \nodata & -/-/- & 0/-/- & 0 \\
2015/06/18 17:24 & 1305 & 1398 &  0.35 & N15E50 & 2015/06/18 16:30 & 2015/06/18 17:36 &          \nodata & 2015/06/18 17:42 & W/-/- & 0/-/- & 0 \\
2015/06/21 02:36 & 1366 & 1740 &  0.97 & N12E16 & 2015/06/21 02:06 & 2015/06/21 02:36 & 2015/06/21 02:24 & 2015/06/21 02:33 & W/-/- & 1/-/- & 1 \\
2015/06/22 18:36 & 1209 & 1573 &  0.60 & N12W08 & 2015/06/22 17:39 & 2015/06/22 18:23 & 2015/06/22 18:05 & 2015/06/22 18:20 & W/-/- & 0/-/- & 0 \\
2015/06/25 08:36 & 1627 & 1805 &  2.15 & N09W42 & 2015/06/25 08:02 & 2015/06/25 08:16 & 2015/06/25 08:16 & 2015/06/25 08:35 & W/-/- & 1/-/- & 1 \\
2015/08/22 07:12 &  547 &  817 &  1.36 & S15E20 & 2015/08/22 06:39 & 2015/08/22 06:49 & 2015/08/22 06:50 & 2015/08/22 07:07 & W/-/- & 0/-/- & 0 \\
2015/09/20 18:12 & 1239 & 1458 &  0.78 & S22W50 & 2015/09/20 17:32 & 2015/09/20 18:03 & 2015/09/20 18:16 & 2015/09/20 18:23 & W/-/- & 0/-/- & 0 \\
2015/11/04 14:48 &  578 &  987 &  0.01 & N09W04 & 2015/11/04 14:08 & 2015/11/05 13:31 & 2015/11/04 13:43 & 2015/11/04 14:07 & W/-/- & 0/-/- & 0 \\
2015/12/16 09:36 &  579 &  937 &  0.43 & S13W04 & 2015/12/16 08:27 & 2015/12/16 09:03 &          \nodata & 2015/12/16 08:45 & W/-/- & 0/-/- & 0 \\
2015/12/28 12:12 & 1212 & 1471 &  0.29 & S23W11 & 2015/12/28 11:20 & 2015/12/28 12:45 &          \nodata & 2015/12/28 11:50 & W/-/- & 0/-/- & 0 \\
2016/01/01 23:24 & 1730 & 1734 &  2.22 & S25W82 & 2016/01/01 23:58 & 2016/01/02 00:11 & 2016/01/01 23:21 & 2016/01/02 00:55 & W/A/- & 1/0/- & 0 \\
2016/02/11 21:17 &  719 & 1174 &  0.43 & N11W07 & 2016/02/11 20:18 & 2016/02/11 21:03 & 2016/02/11 20:35 &          \nodata & -/-/- & 0/0/- & 0 \\
2017/04/18 19:48 &  926 &  932 &  0.32 & N14E77 & 2017/04/18 19:21 & 2017/04/18 20:10 & 2017/04/18 19:49 &          \nodata & -/-/- & 0/1/- & 0 \\
2017/07/14 01:25 & 1200 & 1422 &  0.38 & S06W29 & 2017/07/14 01:07 & 2017/07/14 02:09 &          \nodata & 2017/07/14 01:18 & W/-/- & 1/0/- & 0 \\
2017/09/04 20:36 & 1418 & 1831 &  6.10 & S10W12 & 2017/09/04 20:28 & 2017/09/04 20:33 & 2017/09/04 20:42 & 2017/09/04 20:27 & W/-/- & 1/0/- & 0 \\
2017/09/06 12:24 & 1571 & 1819 &  3.37 & S08W33 & 2017/09/06 11:53 & 2017/09/06 12:02 & 2017/09/06 12:02 & 2017/09/06 12:05 & W/A/- & 1/0/- & 1 \\
2017/09/10 16:00 & 3163 & 3163 &  1.70 & S09W92 & 2017/09/10 15:35 & 2017/09/10 16:06 & 2017/09/10 16:08 & 2017/09/10 16:02 & W/A/- & 1/1/- & 1 \\
\enddata
\tablecomments{The SEP column gives major SEP events observed by GOES and the SGRE column gamma-ray flares with a 'Delayed' component observed by Fermi/LAT \citep{2021ApJS..252...13A}}
\end{deluxetable*}
\end{longrotatetable}

%


\bibliography{SGRE_bibliography}{}
\bibliographystyle{aasjournal}



\end{document}